\documentclass[fleqn,usenatbib,useAMS]{mnras}

\usepackage{graphicx}   % Including figure files
\usepackage{amsmath}    % Advanced maths commands

\usepackage{amssymb}    % Extra maths symbols
\usepackage{multicol}        % Multi-column entries in tables
\usepackage{bm}         % Bold maths symbols, including upright Greek
\usepackage{pdflscape}  % Landscape pages
\usepackage[T1]{fontenc}
\usepackage{ae,aecompl}

\usepackage{newtxtext,newtxmath}
\title{A Study of Warm Dark Matter, the Missing Satellites
Problem, and the UV Luminosity Cut-Off}

\author[B. Hoeneisen]{Bruce Hoeneisen$^{1}$%
\thanks{Contact e-mail: \href{mailto:bhoeneisen@usfq.edu.ec}{bhoeneisen@usfq.edu.ec}}%
\\
$^{1}$Universidad San Francisco de Quito, Quito, Ecuador}

\date{Last updated 2022 November 22; in original form 2022 November 18}

\pubyear{2020}

\begin{document}
\label{firstpage}
\pagerange{\pageref{firstpage}--\pageref{lastpage}}
\maketitle

\begin{abstract}
\noindent
In the warm dark matter	scenario, the Press-Schechter formalism	is
valid only for galaxy masses greater than the ``velocity dispersion cut-off".
In this	work we	extend the predictions to masses below the 
velocity dispersion cut-off, and thereby address the
``Missing Satellites Problem", and the rest-frame ultra-violet 
luminosity cut-off required to not exceed the measured
reionization optical depth. We find agreement between 
predictions and observations of these
two phenomena.
As a by-product, we obtain the empirical Tully-Fisher relation
from first principles.
\end{abstract}


\section{Introduction}

%\citet{author2013} \citep{author2013}
%\citep{jones2015} \citet{smith2014}

Two apparent problems with the cold dark matter	$\Lambda$CDM cosmology
are the ``Missing Satellites Problem", and the need of a 
rest-frame ultra-violet (UV) luminosity cut-off.
The ``Missing Satellites Problem" is the reduced number of observed           
Local Group satellites compared to the number obtained in
$\Lambda$CDM simulations 
\citep{Klypin}.
A UV luminosity cut-off is needed to not exceed the         
reionization optical depth $\tau = 0.054 \pm 0.007$ measured by the
Planck collaboration 
\citep{Planck}
\citep{PDG2022} \citep{Lapi2} \citep{Mason}. 
In the present study we consider warm dark matter as a possible
solution to both problems.

The Press-Schechter formalism, when applied to warm dark matter,
includes the free-streaming cut-off, but not the ``velocity dispersion cut-off",
and is therefore only valid for total (dark matter plus baryon) linear perturbation masses $M$
greater than the velocity dispersion cut-off mass
$M_\textrm{vd}$ (to be explained below). The purpose of the present study is to extend the
Press-Schechter	prediction to $M < M_\textrm{vd}$, and compare this extension
with the ``Missing Satellites Problem", and with the needed UV luminosity cut-off.

\begin{figure}
%\begin{center}
%\vspace*{-4.5cm}
%\scalebox{0.7}
{\includegraphics[width=\columnwidth]{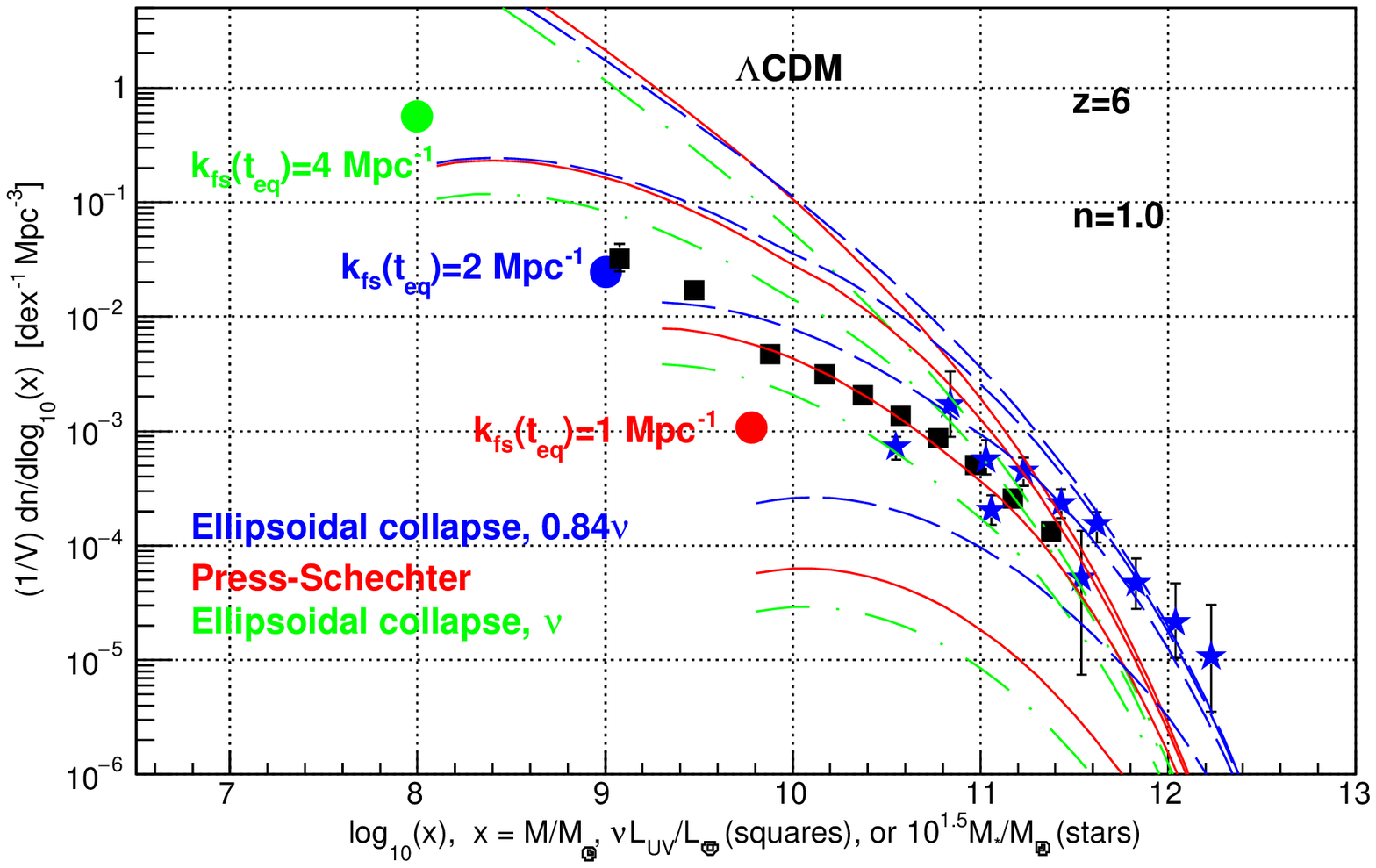}}
\caption{Shown are distributions of $x$, where $x$ is the
observed galaxy stellar mass $M_*/M_\odot$ times $10^{1.5}$ (stars) \citep{Lapi_SMF} \citep{So} \citep{Gr} \citep{Da},
or the observed galaxy UV
luminosity $\nu L_\textrm{UV}/L_\odot$ (squares) \citep{Lapi2} \citep{Bouwens} \citep{Bouwens2021} \citep{McLeod}
(corrected for dust extinction \citep{Lapi2} \citep{Bouwens_27}), or the
predicted linear total (dark matter plus baryon) perturbation mass $M/M_\odot$ (lines),
at redshift $z = 6$.
The Press-Schechter prediction, and its Sheth-Tormen ellipsoidal collapse extensions,
correspond, from top to bottom, to the warm dark matter free-streaming cut-off wavenumbers 
$k_\textrm{fs}(t_\textrm{eq}) = 1000, 4, 2$ and $1$ Mpc$^{-1}$.
The round red, blue and green dots indicate the velocity dispersion cut-offs of the
predictions \citep{first_galaxies} at
$k_\textrm{fs}(t_\textrm{eq}) = 1, 2$ and 4 Mpc$^{-1}$, respectively. 
Presenting three predictions illustrates the uncertainty of the predictions.
}
%%/home/bruce1/first_star/UV_Press_Schechter_LUV.C_bck030722_2_101022
%/home/bruce1/first_star/UV_Press_Schechter_LUV.C_bck030722_2_161122
\label{x_z6}
%\end{center}
\end{figure}

We continue the study of warm dark matter presented in \citet{UVL}.
Our point of departure is Figure 1 of \citet{UVL}.
Here we reproduce the panel corresponding to redshift $z = 6$ in Figure \ref{x_z6}
(with one change: instead of the Gaussian window function in \citet{UVL}, in the present article we use the
sharp-$k$ window function throughout, with mass parameter $c = 1.555$ as explained in \citet{UVL}).
Figure \ref{x_z6} compares distributions, i.e. numbers of galaxies per decade (dex) and per Mpc$^3$, 
of galaxy linear total (dark matter plus baryon) perturbation masses $M$, 
stellar masses $M_*$, and rest-frame ultra-violet (UV)
luminosities $\nu L_{UV}$, with the Press-Schechter prediction \citep{Press-Schechter}, 
and its Sheth-Tormen ellipsoidal collapse extensions  with parameter
$\nu \equiv 1.686/\sigma$ (not to be confused with the frequency above) and $0.84 \nu$
\citep{Sheth_Tormen} \citep{Sheth_Mo_Tormen}. 
The data on $M_*$ is obtained from \citet{Lapi_SMF}, \citet{So}, \citet{Gr}, and \citet{Da}.
The data on $\nu L_{UV}$, where $\nu$ is the frequency corresponding to
wavelength $1550$\AA, is obtained from \citet{Lapi2}, \citet{Bouwens}, \citet{Bouwens2021}, and \citet{McLeod}.
The UV luminosities have been corrected for dust extinction as described in \citet{Lapi2} and \citet{Bouwens_27}.
The predictions depend on the
warm dark matter free-streaming comoving cut-off wavenumber $k_\textrm{fs}(t_\textrm{eq})$, and the comparisons
of predictions with data provide a measurement of $k_\textrm{fs}(t_\textrm{eq})$, see \citet{UVL}
for full details.
In Figure \ref{x_z6} the predictions extend down to the velocity dispersion cut-offs
indicated by red, blue and green dots \citep{UVL} \citep{first_galaxies}.
The purpose of the present study is to extend the predictions to smaller $M_*$ and $\nu L_\textrm{UV}$, and
thereby address the ``Missing Satellites Problem", and the UV luminosities cut-off, respectively.

\section{Velocity Dispersion and Free-Streaming}

To obtain a self-contained article,
we need to define the warm dark matter adiabatic invariant
$v_{h\textrm{rms}}(1)$, and the free-streaming cut-off factor $\tau^2(k)$,
as in \citet{UVL}.
We consider non-relativistic warm dark matter to be a clasical (non-degenerate)
gas of particles, as justified in \citet{Pfenniger} and \citet{wdm_measurements_and_limits}.
Let $v_{h\textrm{rms}}(a)$ be the root-mean-square velocity of non-relativistic 
warm dark matter particles in the early universe at expansion parameter $a$.
As the universe expands it cools, so $v_{h\textrm{rms}}(a)$ decreases in proportion to
$a^{-1}$ (if dark matter collisions, if any, do not excite particle internal
degrees of freedom \citep{dwarf}). Therefore,
\begin{equation}
v_{h\textrm{rms}}(1) = v_{h\textrm{rms}}(a) a =
v_{h\textrm{rms}}(a) \left[ \frac{\Omega_c \rho_\textrm{crit}}{\rho_h(a)} \right]^{1/3},
\label{vhrms1}
\end{equation}
is an adiabatic invariant.
$\rho_h(a) = \Omega_c \rho_\textrm{crit} / a^3$ is the dark matter density.
The warm dark matter velocity dispersion causes free-streaming of dark matter particles
in and out of density minimums and maximums, and so
attenuates the power spectrum 
of relative density perturbations $(\rho(\mathbf{x}) - \bar{\rho})/\bar{\rho}$
of the cold dark matter $\Lambda$CDM cosmology by a factor $\tau^2(k)$. $k$ is
the comoving wavenumber of relative density perturbations.
At the time $t_\textrm{eq}$ of equal radiation and matter densities, $\tau^2(k)$ has the approximate
form \citep{Boyanovsky}
\begin{equation}
\tau^2(k) \approx \exp{\left[-k^2/k^2_\textrm{fs}(t_\textrm{eq})\right]},
\label{tau2}
\end{equation}
where the comoving cut-off wavenumber, due to free-streaming, is \citep{Boyanovsky}
\begin{equation}
k_\textrm{fs}(t_\textrm{eq})
= \frac{1.455}{\sqrt{2}} \sqrt{\frac{4 \pi G \bar{\rho}_h(1) a_\textrm{eq}}{v_{h\textrm{rms}}(1)^2}}.
\label{kfs}
\end{equation}
%$\bar{\rho}_h(1) \equiv \Omega_c \rho_\textrm{crit}$ is the 
%dark matter density at the present time.
After $t_\textrm{eq}$, the Jeans mass decreases as $a^{-3/2}$, so $\tau^2(k)$ develops
a non-linear regenerated ``tail" when the relative density perturbations approach unity \citep{White}.
We will take $\tau^2(k)$, at the time of galaxy formation, to have the form
\begin{eqnarray}
\tau^2(k) & = & \exp{\left( -\frac{k^2}{k^2_\textrm{fs}(t_\textrm{eq})} \right)} \qquad
\textrm{ if } k < k_\textrm{fs}(t_\textrm{eq}), \nonumber \\
& = & \exp{\left( -\frac{k^n}{k^n_\textrm{fs}(t_\textrm{eq})} \right)}\qquad
\textrm{   if } k \ge k_\textrm{fs}(t_\textrm{eq}).
\label{tail}
\end{eqnarray}
The parameter $n$ allows a study of the effect of the non-linear regenerated tail.
If $n = 2$, there is no regenerated tail. Agreement between the data and predictions,
down to the velocity dispersion cut-off dots in Figure \ref{x_z6}, is obtained
with $n$ in the approximate range 1.1 to 0.2 \citep{UVL}.

A comment: In (\ref{tail}) we should have written $k_\textrm{fs}(t_\textrm{gal})$ instead of
$k_\textrm{fs}(t_\textrm{eq})$, where $t_\textrm{gal}$ is the time of galaxy formation.
However, the measurement $k_\textrm{fs}(t_\textrm{gal}) = 2.0^{+0.8}_{-0.5} \textrm{ Mpc}^{-1}$ 
with galaxy UV luminosity
distributions and galaxy stellar mass distributions \citep{UVL}, is in agreement with the measurement
of $k_\textrm{fs}(t_\textrm{eq}) = 1.90 \pm 0.32 \textrm{ Mpc}^{-1}$ with dwarf galaxy rotation 
curves (from the measurement of the adiabatic invariant $v_{h\textrm{rms}}(1) = 0.406 \pm 0.069$ km/s 
in \citet{dwarf}, and Equation (\ref{kfs})).
So we do not distinguish $k_\textrm{fs}(t_\textrm{gal})$ from $k_\textrm{fs}(t_\textrm{eq})$
(until observations require otherwise).

Let us now consider the velocity dispersion cut-off. In the $\Lambda$CDM scenario,
when a spherically symmetric relative density perturbation $(\rho(\mathbf{x}) - \bar{\rho})/\bar{\rho}$ 
reaches 1.686 in the linear approximation, the exact solution diverges, and a galaxy forms.
This is the basis of the Press-Schechter formalism.
The same is true in the warm dark matter scenario if the linear total (dark
matter plus baryon) perturbation mass $M$
exceeds the velocity dispersion cut-off $M_\textrm{vd0}$. For $M < M_\textrm{vd0}$, the
galaxy formation redshift $z$ is delayed by $\Delta z$ due to the velocity dispersion. 
This delay $\Delta z$ is not included in the Press-Schechter formalism. $\Delta z$ is obtained
by numerical integration of the galaxy formation hydro-dynamical equations, see \citet{first_galaxies}.
The velocity dispersion cut-off mass $M_\textrm{vd}$, indicated by the dots in Figure \ref{x_z6}, corresponds,
by definition,
to $\Delta z = 1$. The values of $M_\textrm{vd}$ are presented in \citet{UVL}.
For $M > M_\textrm{vd0} = 10^{0.67} M_\textrm{vd}$ we take $\Delta z = 0$.
For $M < M_\textrm{vd0} = 10^{0.67} M_\textrm{vd}$ we may approximate 
$\Delta z \approx 1.5 \left[\log_{10}(M_\textrm{vd}/M_\odot) + 0.67 - \log_{10}(M/M_\odot)\right]$.
The values of $\log_{10}(M_\textrm{vd}/M_\odot)$ are summarized in Table \ref{vd}.

\begin{table}
\begin{center}
\caption{\label{vd}
The warm dark matter velocity dispersion delays the galaxy formation redshift $z$ by
$\Delta z \approx 1.5 \left[\log_{10}(M_\textrm{vd}/M_\odot) + 0.67 - \log_{10}(M/M_\odot)\right]$
if $M < M_\textrm{vd0} = 10^{0.67} M_\textrm{vd}$.
The values of $\log_{10}(M_\textrm{vd}/M_\odot)$ are presented as a function
of the galaxy formation redshift $z$, and the adiabatic invariant $v_{h\textrm{rms}}(1)$.
$M_\textrm{vd}$ is obtained from numerical integrations
of galaxy formation hydro-dynamical equations \citep{first_galaxies}.
Also shown is $k_\textrm{fs}(t_\textrm{eq})$ form (\ref{kfs}).
By definition, at $M = M_\textrm{vd}$, $\Delta z = 1.0$.
}
\begin{tabular}{lllr}
\hline
\hline
$z$ & $v_{h\textrm{rms}}(1)$ & $k_\textrm{fs}(t_\textrm{eq})$ & $\log_{10}(M_\textrm{vd}/M_\odot)$ \\
%    & [km/s]                             & [Mpc$^{-1}$]                   & \\
\hline
4 & 0.75 km/s & 1 Mpc$^{-1}$  & 9.3  \\
4 & 0.49 km/s & 1.53 Mpc$^{-1}$ & 8.5  \\
4 & 0.37 km/s & 2 Mpc$^{-1}$  & 8.3  \\
4 & 0.19 km/s & 4 Mpc$^{-1}$  & 7.5  \\
6 & 0.75 km/s & 1 Mpc$^{-1}$  & 9.8  \\
6 & 0.49 km/s & 1.53 Mpc$^{-1}$ & 9.3  \\
6 & 0.37 km/s & 2 Mpc$^{-1}$  & 9.0  \\
6 & 0.19 km/s & 4 Mpc$^{-1}$  & 8.0  \\
8 & 0.75 km/s & 1 Mpc$^{-1}$  & 10.3  \\
8 & 0.49 km/s & 1.53 Mpc$^{-1}$ & 9.6  \\
8 & 0.37 km/s & 2 Mpc$^{-1}$  & 9.2  \\
8 & 0.19 km/s & 4 Mpc$^{-1}$  & 8.2  \\
\hline
\hline
\end{tabular}
\end{center}
\end{table}

\section{Extending the Predictions to $M < M_\textrm{vd0}$}

\begin{figure}
%\begin{center}
%\vspace*{-4.5cm}
%\scalebox{0.325}
{\includegraphics[width=\columnwidth]{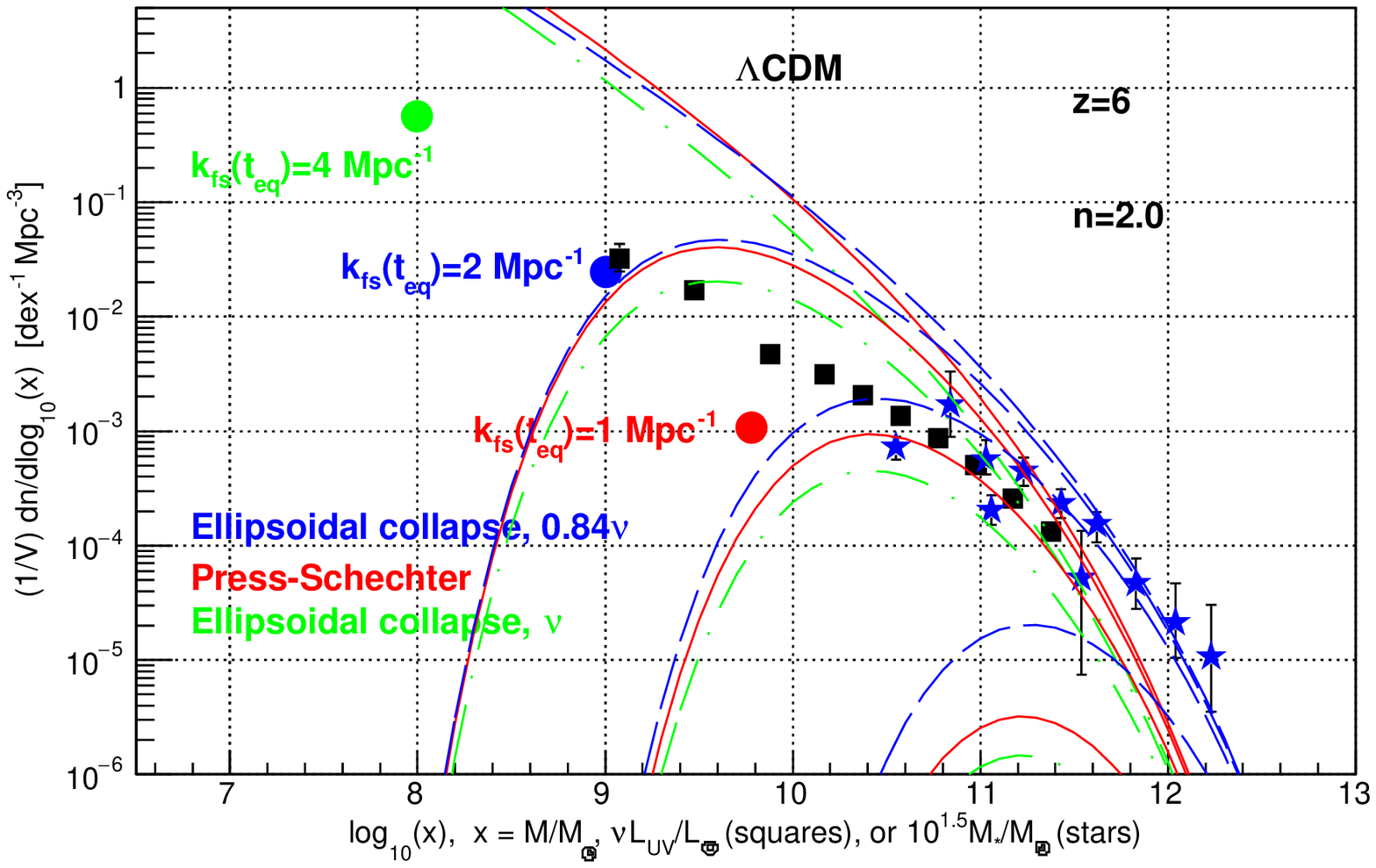}} \\
%\scalebox{0.325}
{\includegraphics[width=\columnwidth]{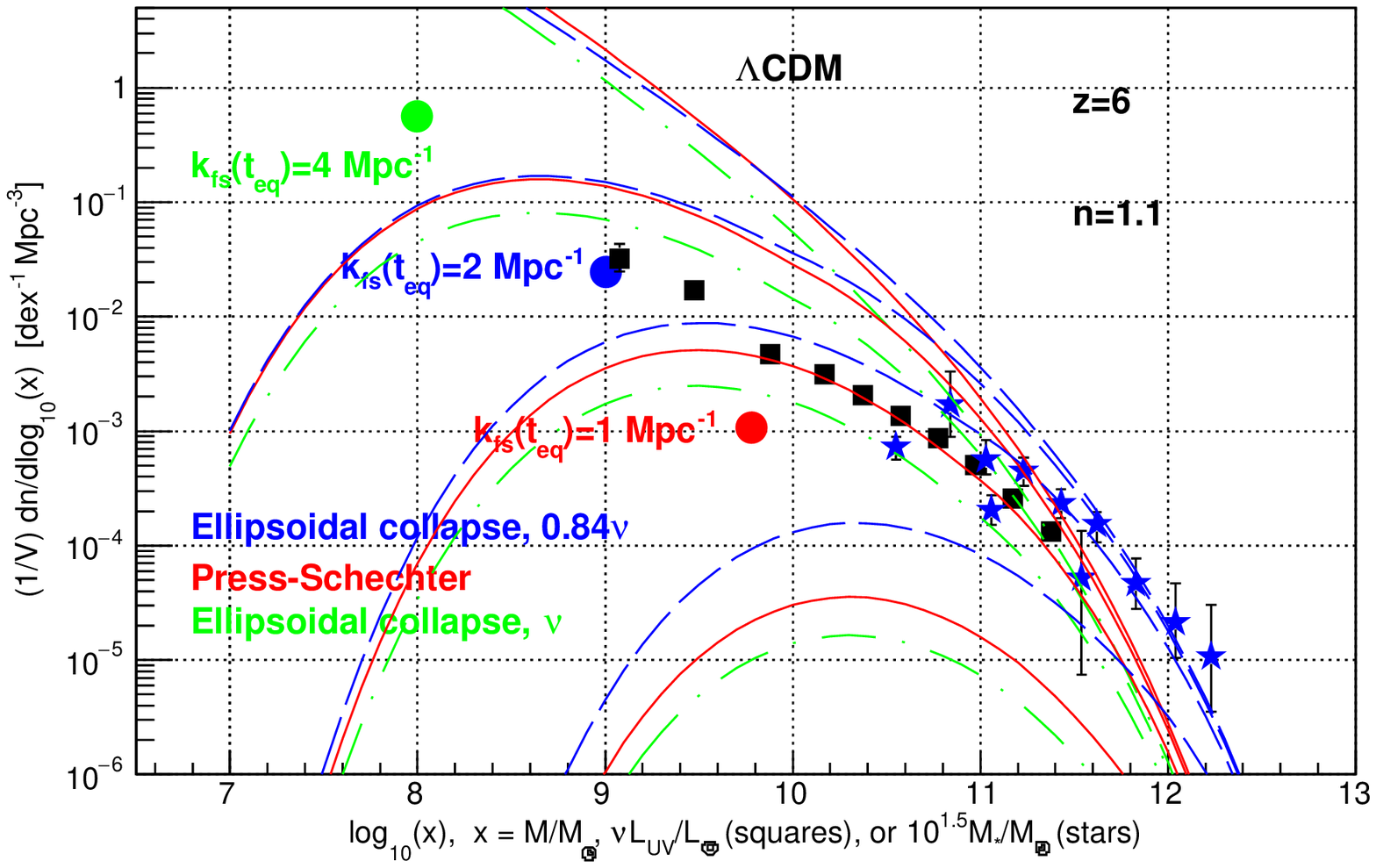}} \\
%\scalebox{0.325}
{\includegraphics[width=\columnwidth]{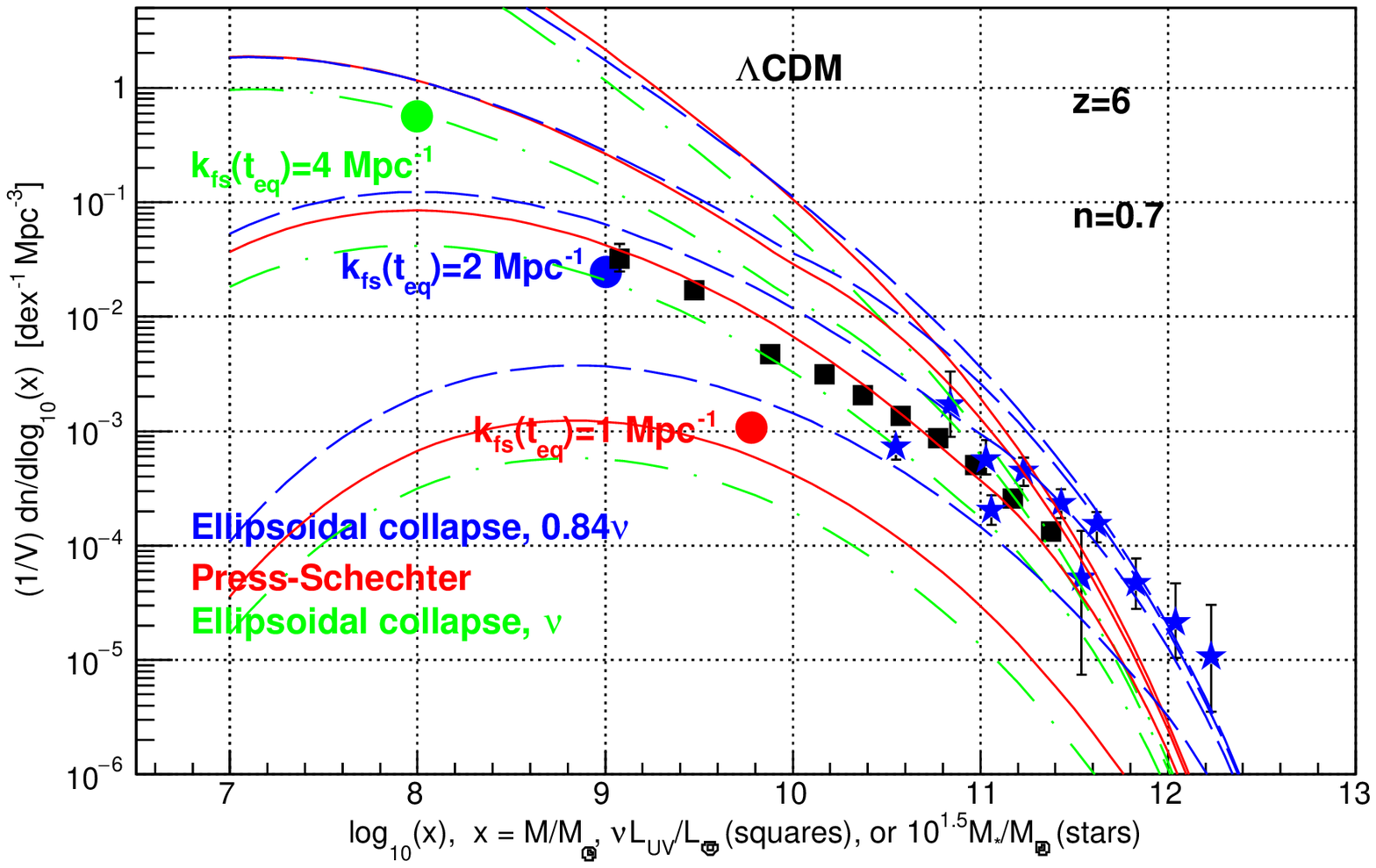}} \\
%\scalebox{0.325}
{\includegraphics[width=\columnwidth]{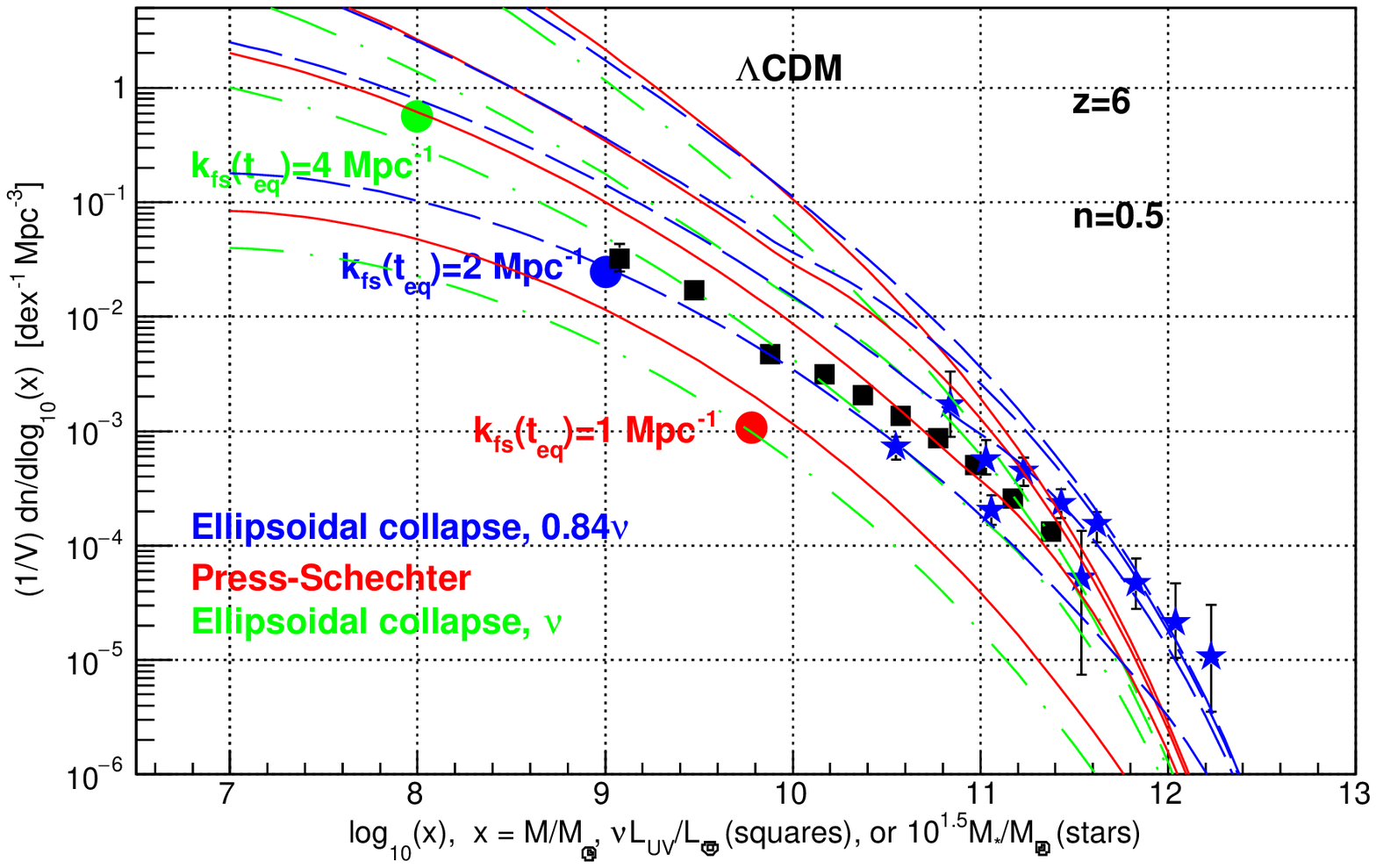}}
\caption{ Same as Figure \ref{x_z6}, i.e. $z = 6$, but the predictions are extended
to $M < M_{\textrm{vd}0}$ with the free-streaming cut-off
with $\tau^2(k)$ with a tail with $n = 2.0, 1.1, 0.7$ or 0.5,
\textit{without} the velocity dispersion cut-off.
}
%~/first_star/UV_Press_Schechter.C_bck121022
\label{z6_Dzz0}
%\end{center}
\end{figure}

\begin{figure}
%\begin{center}
%\vspace*{-4.5cm}
%\scalebox{0.325}
%{\includegraphics[width=\columnwidth]{UV_Press_Schechter_311022_z8_N2.0_Dzz.eps}} \\
%\scalebox{0.325}
{\includegraphics[width=\columnwidth]{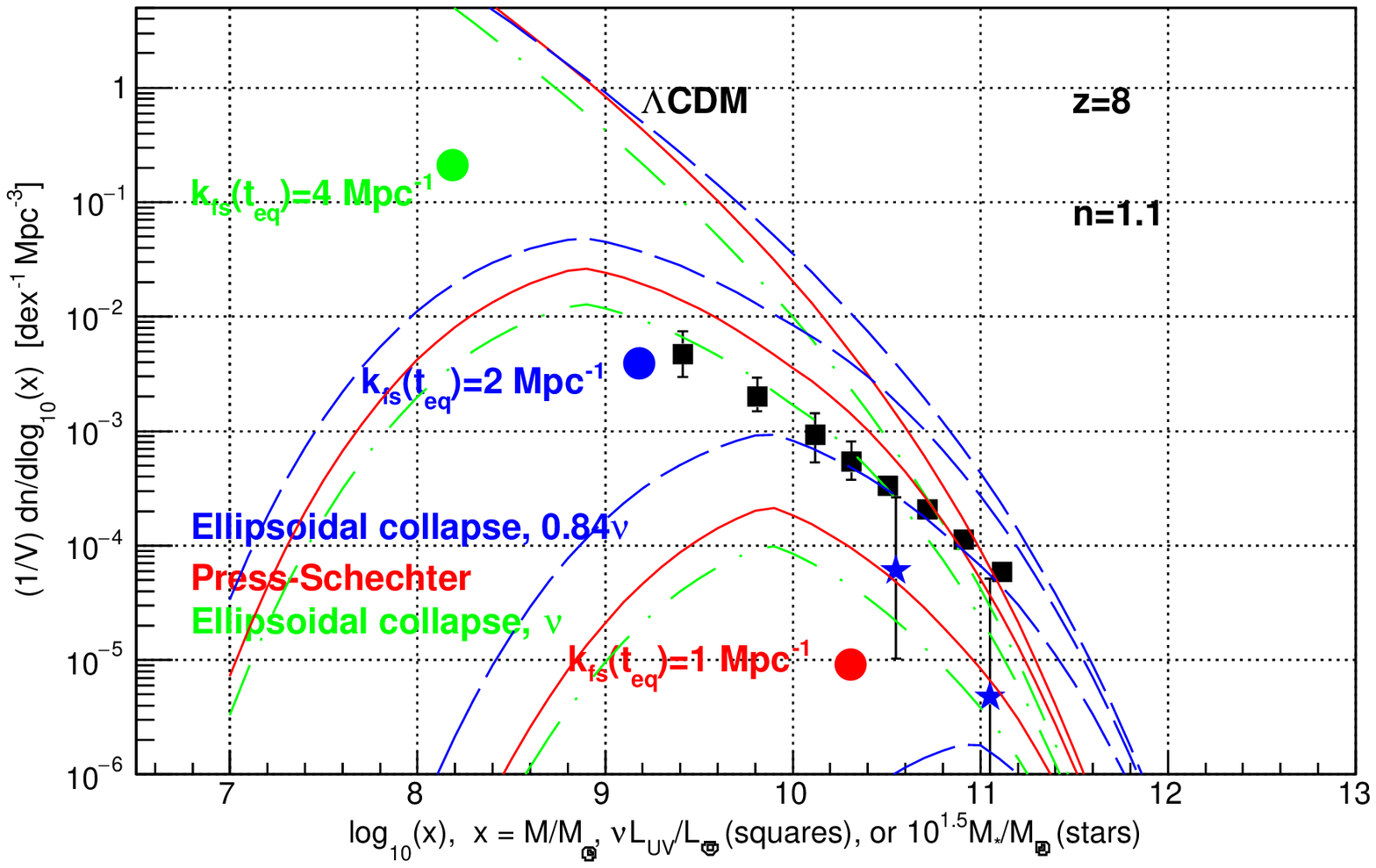}} \\
%\scalebox{0.325}
{\includegraphics[width=\columnwidth]{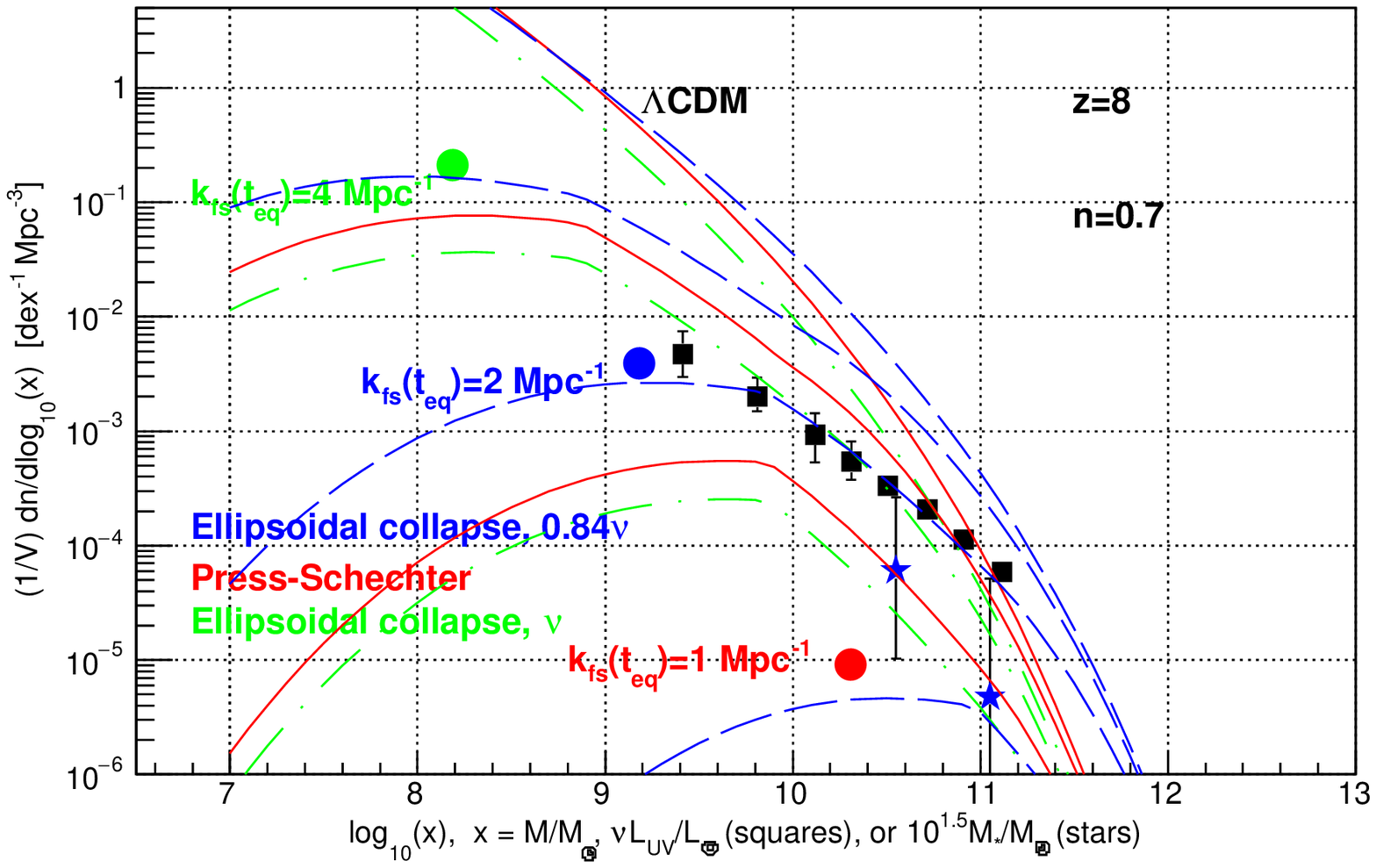}} \\
%\scalebox{0.325}
{\includegraphics[width=\columnwidth]{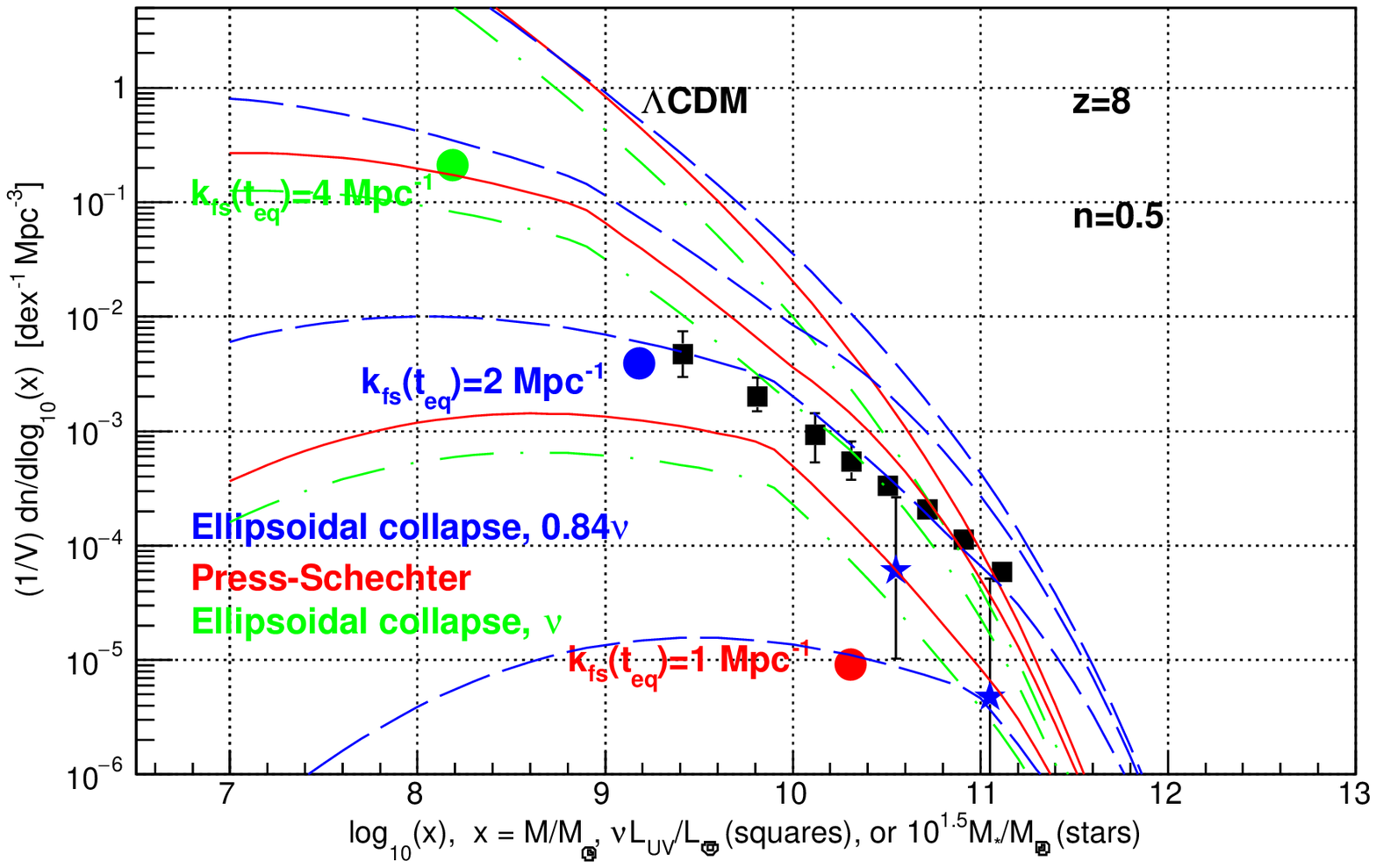}} \\
%\scalebox{0.325}
{\includegraphics[width=\columnwidth]{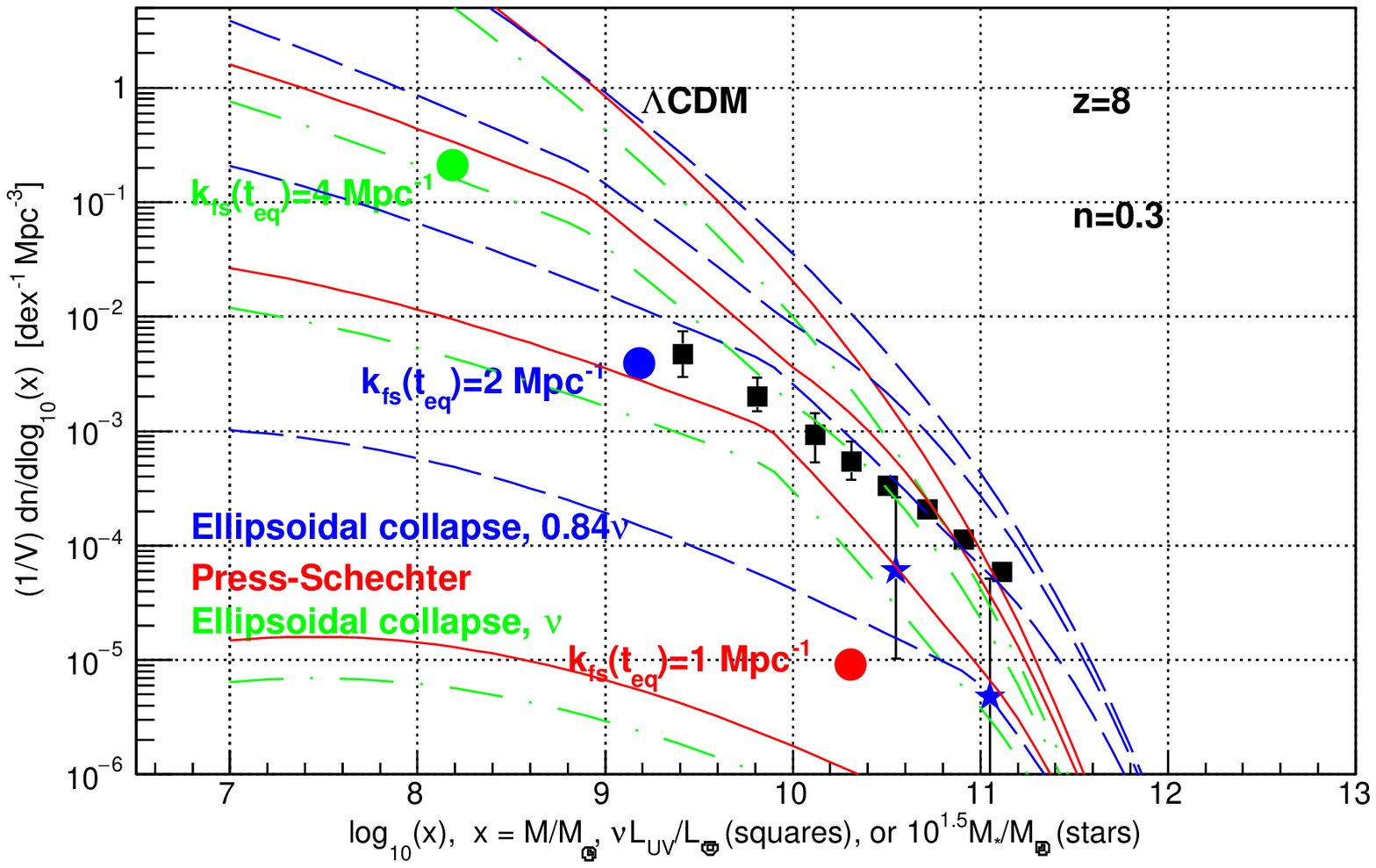}} \\
%\scalebox{0.325}
%{\includegraphics[width=\columnwidth]{UV_Press_Schechter_311022_z8_N0.1_Dzz.eps}} \\
\caption{Predictions for $z = 8$, and (from top to bottom)
$k_\textrm{fs}(t_\textrm{eq}) = 1000, 4, 2, 1 \textrm{ Mpc}^{-1}$, are extended
to $M < M_{\textrm{vd}0}$ with the free-streaming cut-off
with $\tau^2(k)$ with a tail with $n = 1.1, 0.7, 0.5$,or 0.3,
and with the velocity dispersion cut-off.
Agreement between predictions and observations are obtained with
$k_\textrm{fs}(t_\textrm{eq}) \approx 2 \textrm{ Mpc}^{-1}$, 
and $0.8 \gtrsim n \gtrsim 0.3$.
}
%~/first_star/UV_Press_Schechter.C_bck121022
%~/first_star/UV_Press_Schechter.C_bck311022
\label{z8_Dzz}
%\end{center}
\end{figure}

\begin{figure}
%\begin{center}
%\vspace*{-4.5cm}
%\scalebox{0.325}
%{\includegraphics[width=\columnwidth]{UV_Press_Schechter_311022_z6_N2.0_Dzz.eps}} \\
%\scalebox{0.325}
{\includegraphics[width=\columnwidth]{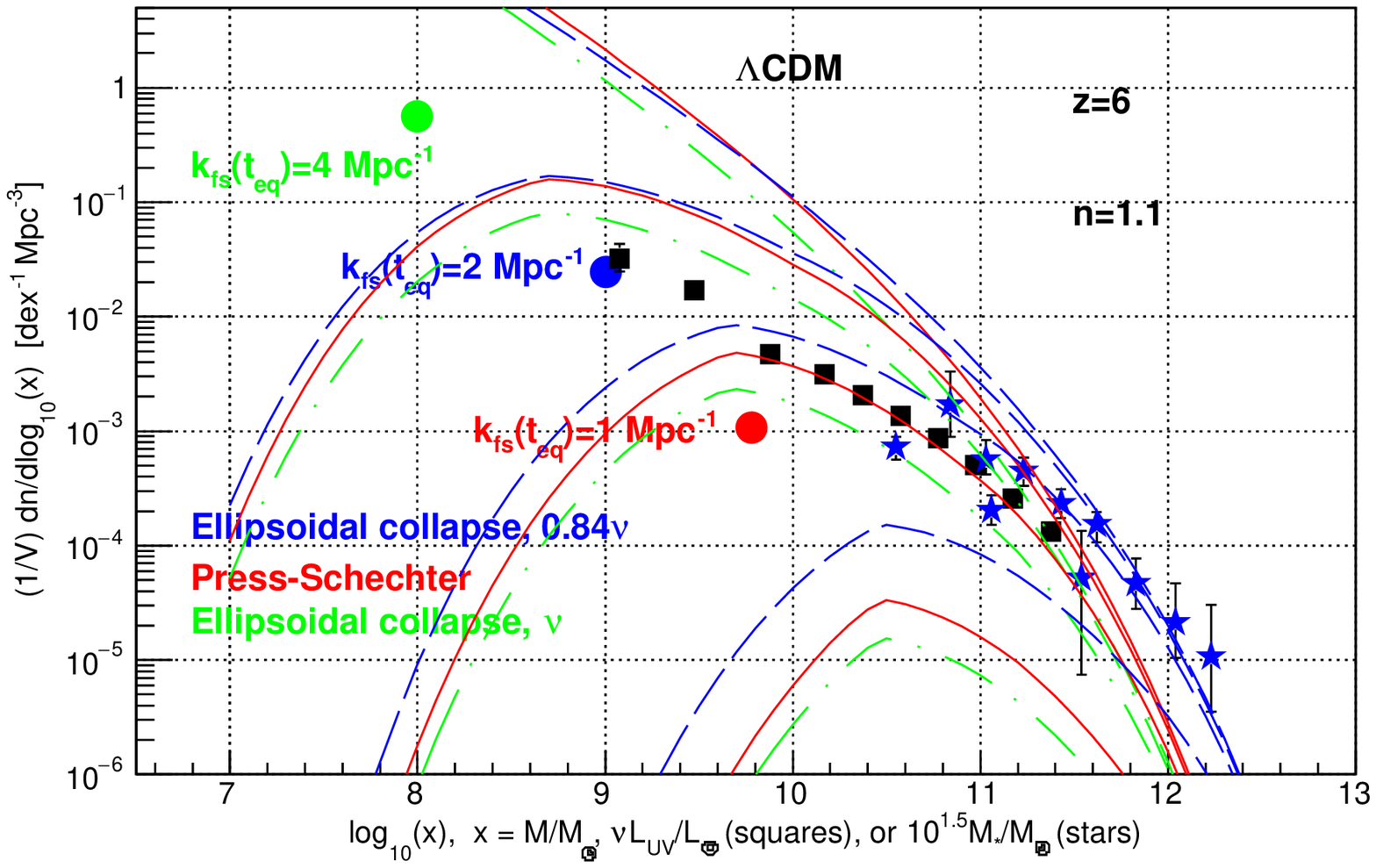}} \\
%\scalebox{0.325}
{\includegraphics[width=\columnwidth]{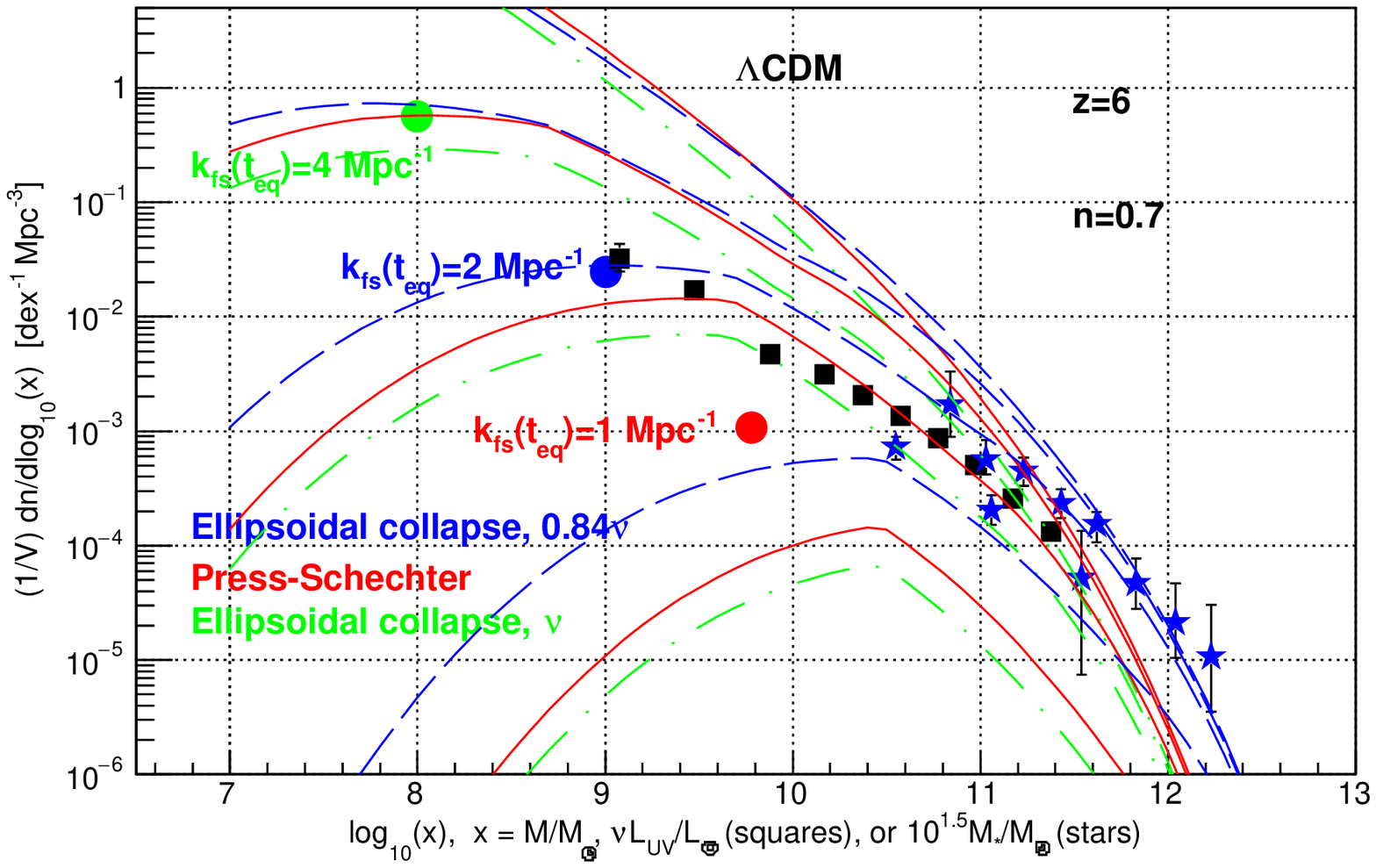}} \\
%\scalebox{0.325}
{\includegraphics[width=\columnwidth]{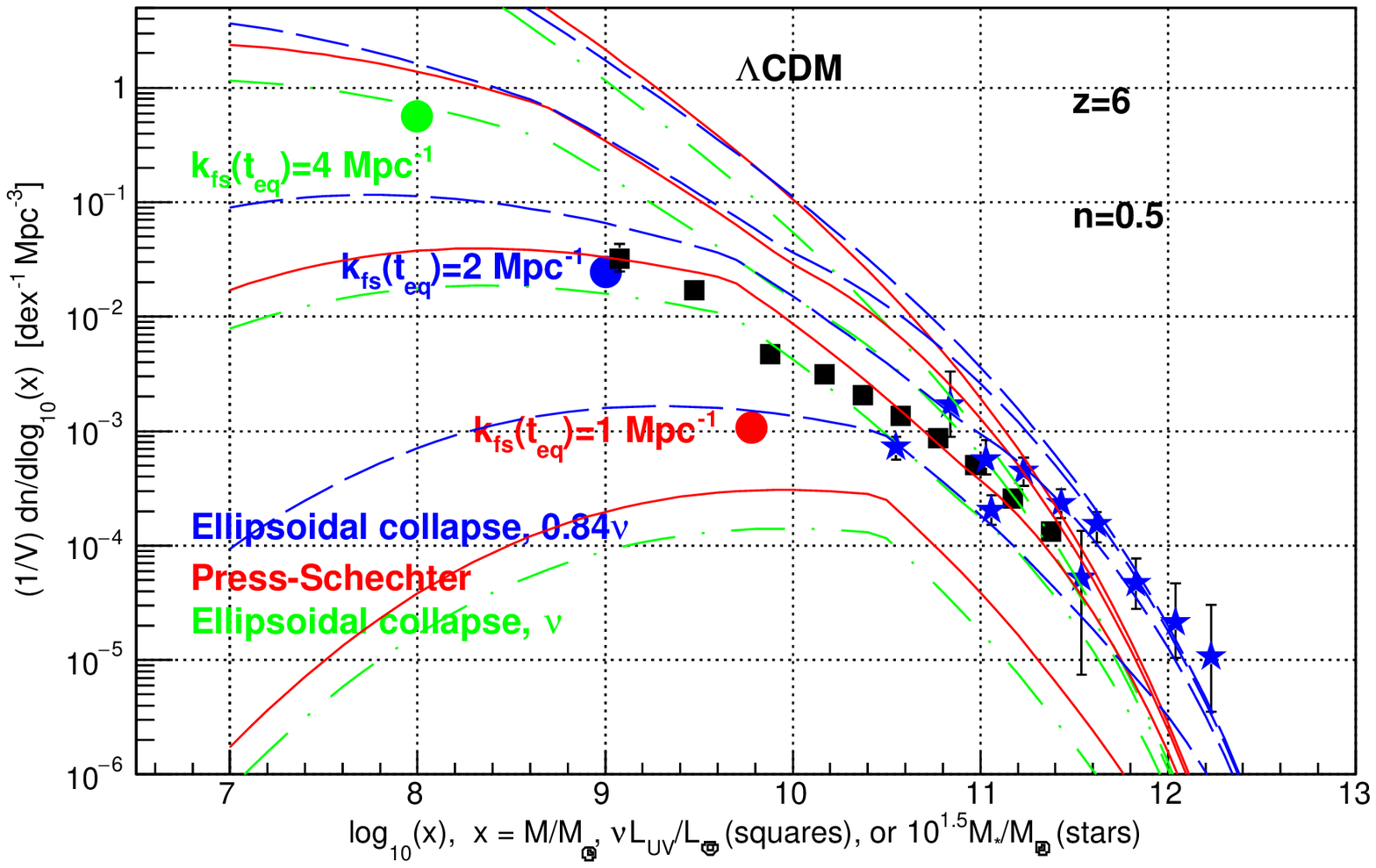}} \\
%\scalebox{0.325}
{\includegraphics[width=\columnwidth]{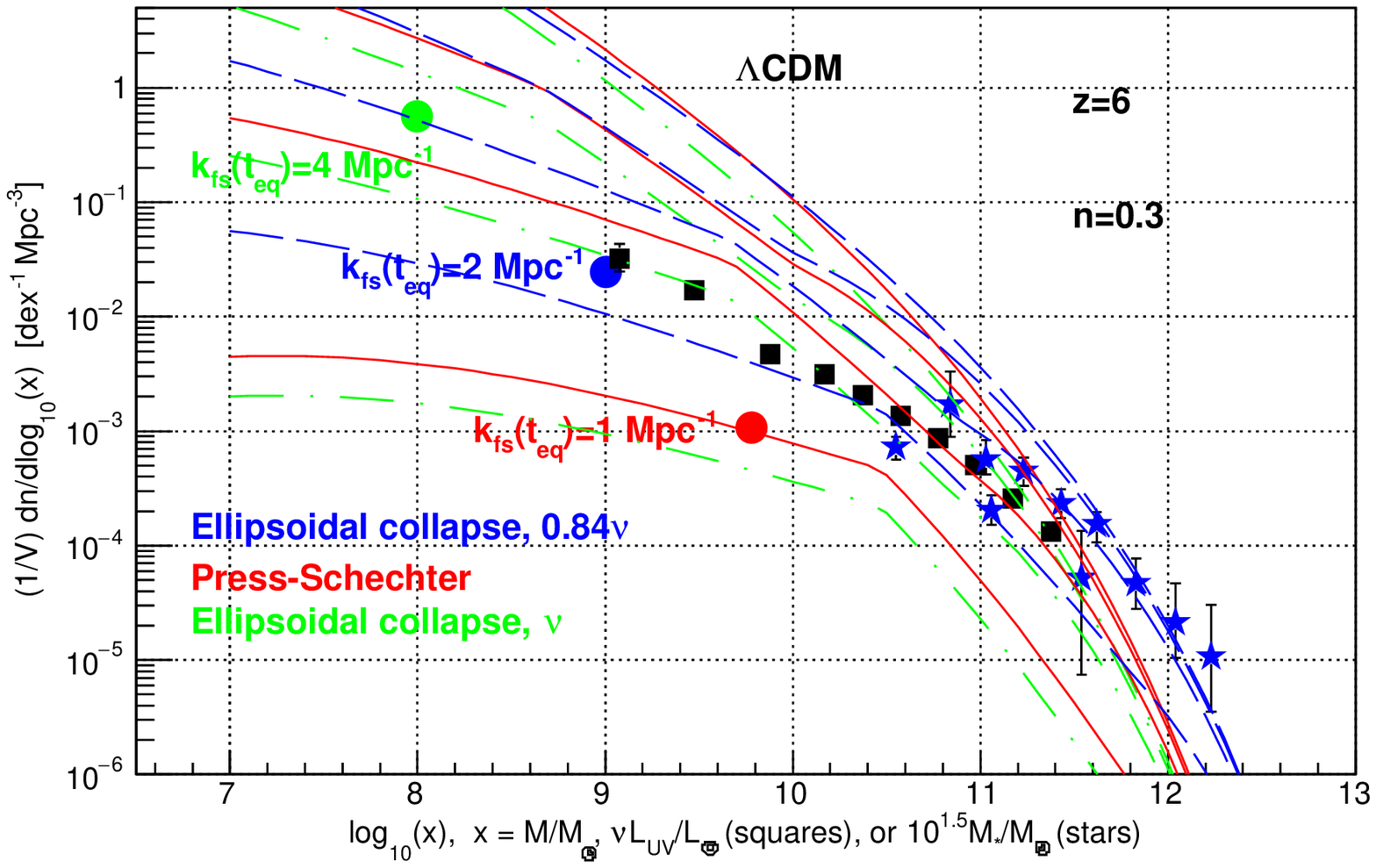}} \\
%\scalebox{0.325}
%{\includegraphics[width=\columnwidth]{UV_Press_Schechter_311022_z6_N0.1_Dzz.eps}} \\
\caption{Predictions for $z = 6$, and (from top to bottom)
$k_\textrm{fs}(t_\textrm{eq}) = 1000, 4, 2, 1 \textrm{ Mpc}^{-1}$, are extended
to $M < M_{\textrm{vd}0}$ with the free-streaming cut-off
with $\tau^2(k)$ with a tail with $n = 1.1, 0.7, 0.5$, or 0.3,
and with the velocity dispersion cut-off.
Agreement between predictions and observations are obtained with
$k_\textrm{fs}(t_\textrm{eq}) \approx 2	\textrm{ Mpc}^{-1}$, 
and $0.8 \gtrsim n \gtrsim 0.3$.
}
%~/first_star/UV_Press_Schechter.C_bck121022
%~/first_star/UV_Press_Schechter.C_bck311022
\label{z6_Dzz}
%\end{center}
\end{figure}

\begin{figure}
%\begin{center}
%\vspace*{-4.5cm}
%\scalebox{0.325}
%{\includegraphics[width=\columnwidth]{UV_Press_Schechter_311022_z4_N2.0_Dzz.eps}} \\
%\scalebox{0.325}
{\includegraphics[width=\columnwidth]{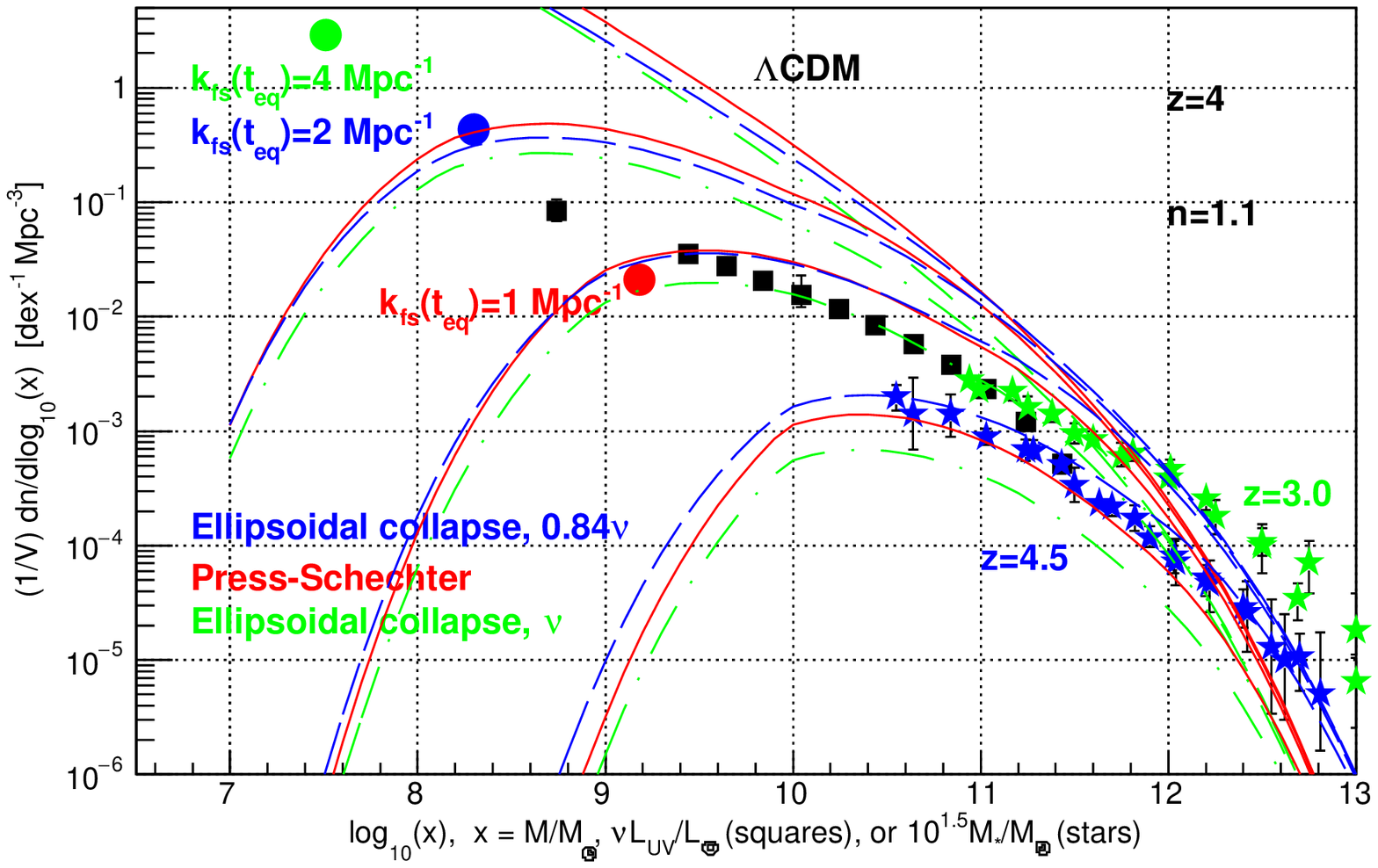}} \\
%\scalebox{0.325}
{\includegraphics[width=\columnwidth]{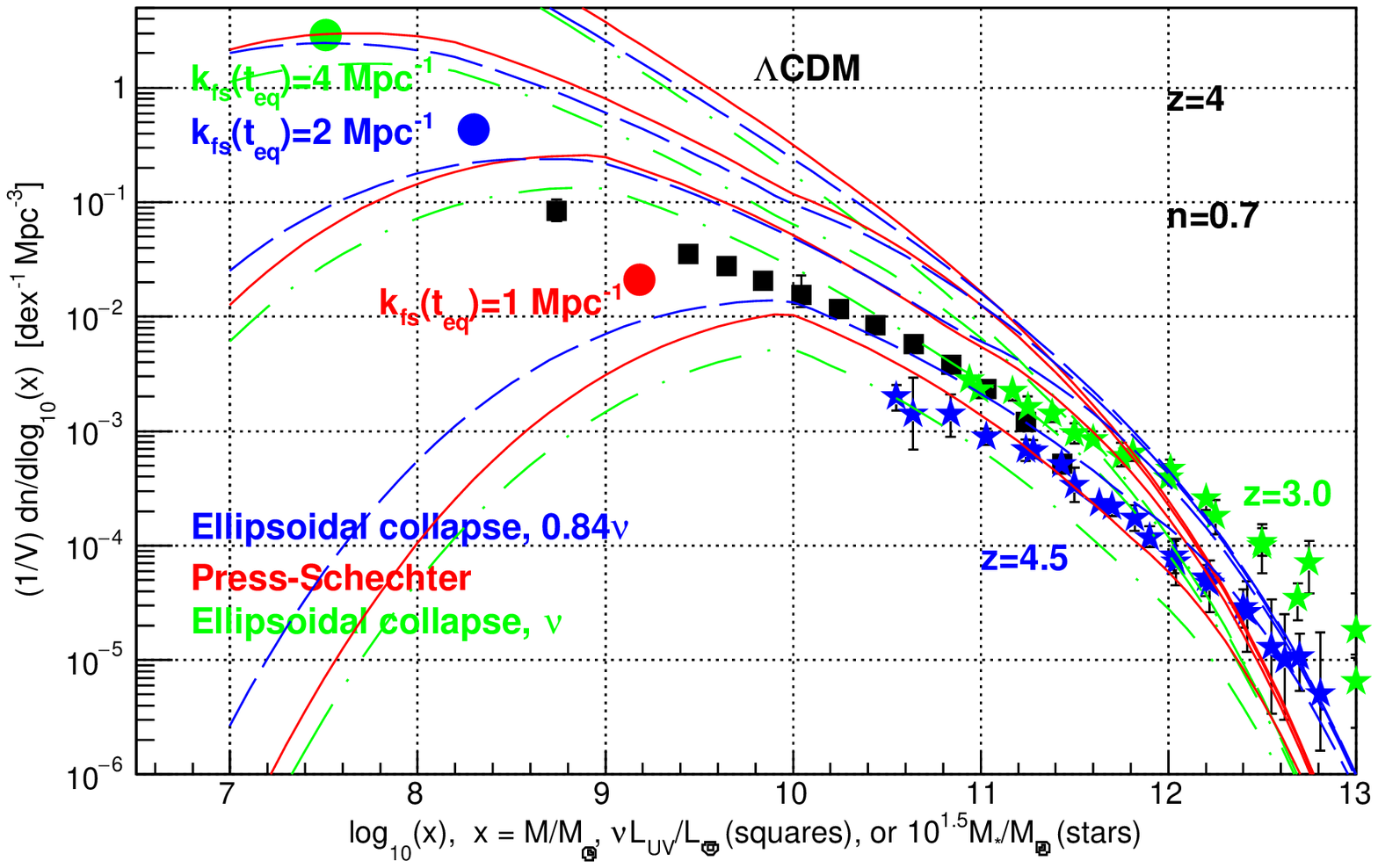}} \\
%\scalebox{0.325}
{\includegraphics[width=\columnwidth]{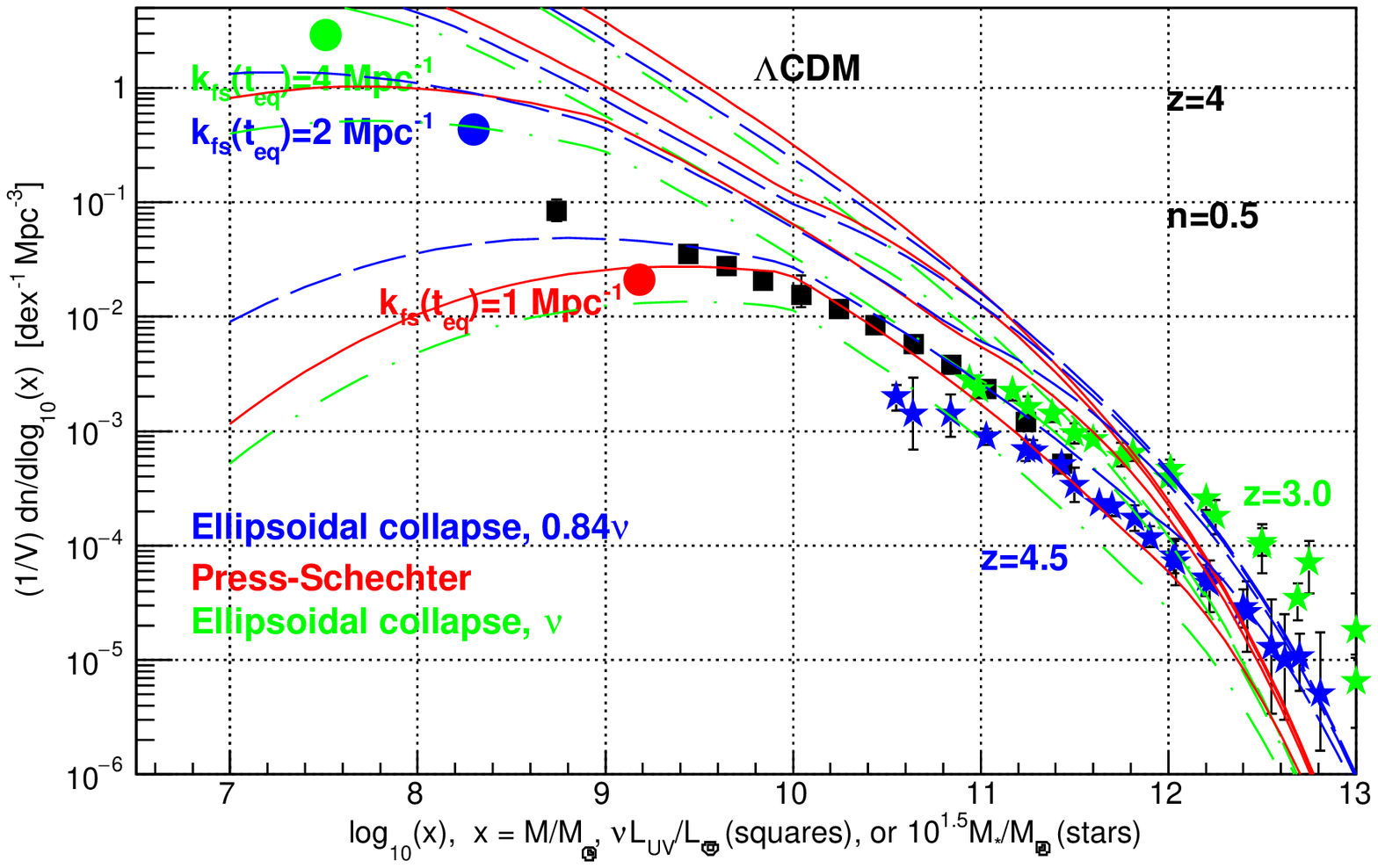}} \\
%\scalebox{0.325}
{\includegraphics[width=\columnwidth]{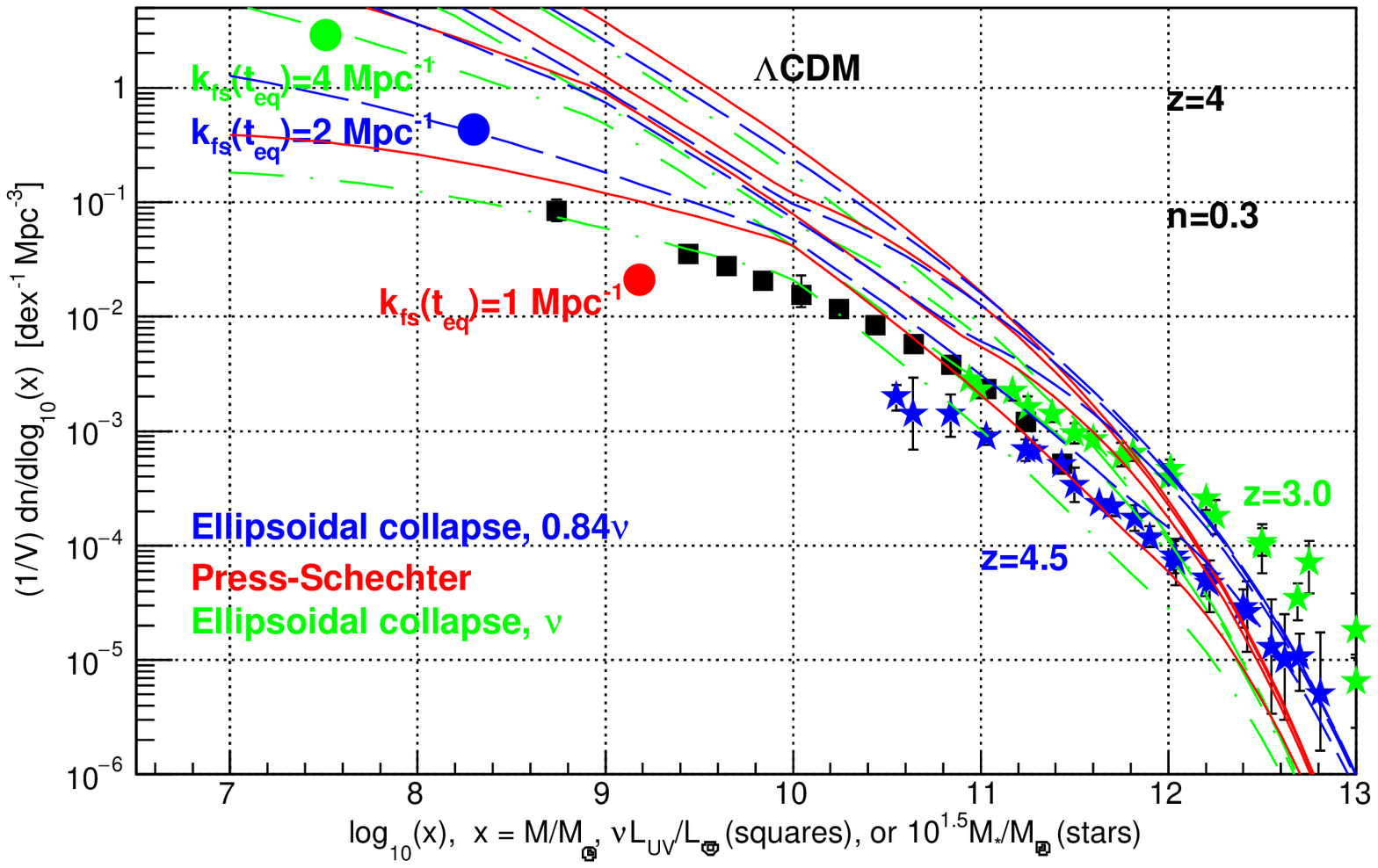}} \\
%\scalebox{0.325}
%{\includegraphics[width=\columnwidth]{UV_Press_Schechter_121022_z4_N0.1_Dzz.eps}} \\
\caption{Predictions for $z = 4$, and (from top	to bottom) 
$k_\textrm{fs}(t_\textrm{eq}) = 1000, 4, 2, 1 \textrm{ Mpc}^{-1}$, are extended
to $M < M_{\textrm{vd}0}$ with the free-streaming cut-off
with $\tau^2(k)$ with a tail with $n = 1.1, 0.7, 0.5$, or 0.3,
and with the velocity dispersion cut-off.
Agreement between predictions and observations are obtained with
$k_\textrm{fs}(t_\textrm{eq}) \approx 1.5 \textrm{ Mpc}^{-1}$ 
and $n \approx 0.7$, or
$k_\textrm{fs}(t_\textrm{eq}) \approx 1 \textrm{ Mpc}^{-1}$
and $n \approx 0.5$.
}
%~/first_star/UV_Press_Schechter.C_bck121022
%~/first_star/UV_Press_Schechter.C_bck311022
\label{z4_Dzz}
%\end{center}
\end{figure}

The Press-Schechter prediction, and its extensions, are based on
the variance $\sigma^2(M, z, k_\textrm{fs}(t_\textrm{eq}),n)$
of the linear relative density perturbation
$(\rho(\mathbf{x}) - \bar{\rho})/\bar{\rho}$ at the total (dark matter plus baryon)
mass scale $M$ \citep{UVL} \citep{Weinberg}.
This variance depends on the redshift $z$ of galaxy formation,
and on the parameters $k_\textrm{fs}(t_\textrm{eq})$ and $n$ of the
free-streaming cut-off factor $\tau^2(k)$ of (\ref{tail}).
Comparison of predictions and data for $M > M_\textrm{vd0}$ obtain
a measurement of $k_\textrm{fs}(t_\textrm{eq})$, see Figure \ref{x_z6}, and \citet{UVL}.
The extension of the predictions
to $M < M_\textrm{vd0}$ depends on two cut-offs:
the free-streaming cut-off (through the parameters $k_\textrm{fs}(t_\textrm{eq}) = 2.0^{+0.8}_{-0.5} \textrm{ Mpc}^{-1}$, 
that is already fixed by the measurements in \citet{UVL}, and $n$), and the
velocity dispersion cut-off.
We illustrate the effect of $n$ in Figure \ref{z6_Dzz0}, 
without applying the velocity dispersion cut-off yet.
The velocity dispersion cut-off
is implemented by replacing $\sigma^2(M, z, k_\textrm{fs}(t_\textrm{eq}),n)$
by $\sigma^2(M, z + \Delta z, k_\textrm{fs}(t_\textrm{eq}),n)$, with $\Delta z$
obtained from Table \ref{vd}.
We illustrate the effect of both $n$, and the velocity dispersion cut-off, in
Figures \ref{z8_Dzz}, \ref{z6_Dzz}, and \ref{z4_Dzz}, for galaxy
formation at $z = 8, 6$, and 4, respectively.

\section{The Relation Between $M$ and $V_\textrm{flat}$}

\begin{table}
\begin{center}
{\small
\caption{\label{V_flat}
The galaxy flat rotation velocity $V_\textrm{flat}$ [km/s] is presented as a function
of the adiabatic invariant $v_{h\textrm{rms}}(1)$, and the galaxy 
formation redshift $z$, for linear perturbations of total (dark matter plus baryon)
mass $M = 2 \times 10^{10} M_\odot$.
Also shown is the free-streaming cut-off wavenumber $k_\textrm{fs}(t_\textrm{eq})$ from (\ref{kfs}).
$V_\textrm{flat}$ is obtained from numerical integration of galaxy formation
hydro-dynamical equations \citep{first_galaxies}.
}
\begin{tabular}{lccccr}
\hline
\hline
%$v_{h\textrm{rms}}(1)$ [km/s] & 790 & 490 & 406 & 200 \\ 
%$k_\textrm{fs}(t_\textrm{eq})$ [Mpc$^{-1}$] & 1 & 1.66 & 2 & 4 \\
$v_{h\textrm{rms}}(1)$ [m/s] & 750 & 490 & 370 & 190 & 0.75 \\ 
$k_\textrm{fs}(t_\textrm{eq})$ [Mpc$^{-1}$] & 1 & 1.53 & 2 & 4 & 1000 \\
$z$ & & & & \\
\hline
4  & 38 & 33 & 36 & 37 & 34 \\
5  & 41 & 35 & 37 & 37 & 37 \\
6  & 47 & 44 & 40 & 49 & 42 \\
8  & 45 & 42 & 45 & 46 & 49 \\
10 & 51 & 49 & 47 & 53 & 51 \\
\hline
\hline
\end{tabular}
}
\end{center}
\end{table}

\begin{table}
\begin{center}
{\small
\caption{\label{M_V_flat}
Shown are linear perturbation total (dark matter plus baryon) masses
$M$, and the corresponding flat rotation velocities $V_\textrm{flat}$.
These relations are approximately valid
for galaxy formation at redshift $z$ between 6 and 10, and
free-streaming cut-off wavenumber $k_\textrm{fs}(t_\textrm{eq})$ between 1 and 1000 Mpc$^{-1}$.
}
\begin{tabular}{lcccr}
\hline
\hline
$M$ & $V_\textrm{flat}$ \\
\hline
$3 \times 10^{12} M_\odot$ & 255 km/s \\ 
$1 \times 10^{12} M_\odot$ & 188 km/s \\ 
$1 \times 10^{11} M_\odot$ & 82 km/s \\ 
$5 \times 10^{10} M_\odot$ & 71 km/s \\ 
$2 \times 10^{10} M_\odot$ & 45 km/s \\ %32*sqrt(2)=45 
$1 \times 10^{10} M_\odot$ & 37 km/s \\ 
$5 \times 10^{9} M_\odot$ & 28 km/s \\ 
$3 \times 10^{8} M_\odot$ & 9 km/s \\ 
\hline
\hline
\end{tabular}
}
\end{center}
\end{table}

The linear perturbation total (dark matter plus baryon) mass scale $M$, of the
Press-Schechter formalism, can not be measured directly. We find that
the flat portion of the rotation velocity of test particles in spiral galaxies, $V_\textrm{flat}$,
can be used as an approximate proxy for $M$.

\begin{figure}
%\begin{center}
%\vspace*{-4.5cm}
%\scalebox{0.7}
{\includegraphics[width=\columnwidth]{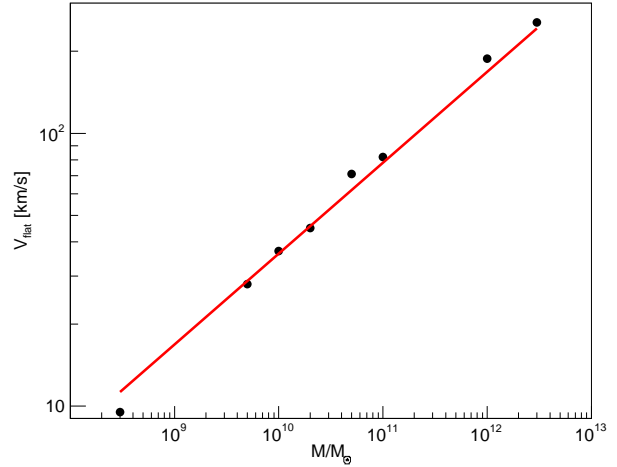}}
\caption{Presented is $V_\textrm{flat}$, from Table \ref{M_V_flat}, as a function of the linear
total (dark matter plus baryon) perturbation mass $M$,
valid for galaxy formation at redshift $z$ between 6 and 10, and
free-streaming cut-off wavenumber $k_\textrm{fs}(t_\textrm{eq})$ between 1 and 1000 Mpc$^{-1}$.
The line is $M/M_\odot = 2.1 \times 10^8 (V_\textrm{flat}/10 \textrm{ km/s})^3$.
}
%~/missing_satellite/M_Vflat.C_bck161022
\label{M_V_flat_fig}
%\end{center}
\end{figure}

Given $M$, the galaxy formation redshift $z$, and $v_{h\textrm{rms}}(1)$, it is possible to obtain $V_\textrm{flat}$
by numerical integration of the galaxy formation hydro-dynamical equations \citep{first_galaxies}.
Results for $M = 2 \times 10^{10} M_\odot$ are presented in Table \ref{V_flat}.
We note that for galaxy formation at redshift $z$ between 6 and 10, and
free-streaming cut-off wavenumber $k_\textrm{fs}(t_\textrm{eq})$ between 1 and 1000 Mpc$^{-1}$,
we may approximate $V_\textrm{flat} \approx 45$ km/s. 
Similarly, for several masses $M$, the corresponding 
rotation velocities $V_\textrm{flat}$ are summarized
in Table \ref{M_V_flat}. The data in Table \ref{M_V_flat} can be fit by the relation
\begin{equation}
\frac{M}{M_\odot} \approx 2.1 \times 10^8 \left(\frac{V_\textrm{flat}}{10 \textrm{ km/s}}\right)^3,
\label{Vflat_vs_M}
\end{equation}
as shown in Figure \ref{M_V_flat_fig}.
This becomes the Tully-Fisher relation, once $M/M_\odot$ is replaced by
$\approx 10^{1.5} M_*/M_\odot \approx 10^{1.5} L_*/L_\odot$, see Figure \ref{x_z6}:
\begin{equation}
\frac{L_*}{L_\odot} \approx 2.4 \times 10^{10} h^{-2} \left(\frac{V_\textrm{flat}}{200 \textrm{ km/s}}\right)^3,
\label{Tully_Fisher}
\end{equation}
with $h = 0.674$. $L_*$ is the stellar luminosity.
It is very satisfactory to obtain quantitatively the empirical
Tully-Fisher relation from the galaxy formation hydro-dynamical equations \citep{first_galaxies}, i.e.
from first principles.

\section{The ``Missing Satellites Problem"}

The ``Missing Satellites Problem" is described in \citet{Klypin}.
The approximate number of observed satellites within $200 h^{-1}$ kpc of the Local Group, per Mpc$^3$,
with $V_\textrm{flat} > V$ is \citep{Klypin}
\begin{equation}
n_\textrm{obs}(V_\textrm{flat} > V) \approx 385 h^3 \left(\frac{10 \textrm{ km/s}}{V}\right)^{1.3} \textrm{ Mpc}^{-3},
\label{n_observed}
\end{equation}
while the corresponding number in the $\Lambda$CDM simulations is \citep{Klypin}
\begin{equation}
n_\textrm{sim}(V_\textrm{flat} > V) \approx 5000 h^3 \left(\frac{10 \textrm{ km/s}}{V}\right)^{2.75} \textrm{ Mpc}^{-3}.             
\label{n_simulated}
\end{equation}
Using (\ref{Vflat_vs_M}), we find that the appropriate ratio to compare the $\Lambda$CDM 
simulation with	observations is
\begin{equation}
\frac{dn_\textrm{sim}/dV_\textrm{flat}}{dn_\textrm{obs}/dV_\textrm{flat}} \approx 
\frac{5000 \cdot 2.75}{385 \cdot 1.3} \left(\frac{10 \textrm{ km/s}}{V_\textrm{flat}}\right)^{1.45}.
\label{dnsim_dnobs_200}
\end{equation}
Similarly, for satellites within $400 h^{-1}$ kpc of the Local Group, the ratio is
\begin{equation}
\frac{dn_\textrm{sim}/dV_\textrm{flat}}{dn_\textrm{obs}/dV_\textrm{flat}} \approx 
\frac{1200 \cdot 2.75}{55 \cdot 1.4} \left(\frac{10 \textrm{ km/s}}{V_\textrm{flat}}\right)^{1.35}.
\label{dnsim_dnobs_400}
\end{equation}
We take agreement between observations and simulations at $V_\textrm{flat} \approx 70$ km/s,
corresponding to $M \approx 10^{10.9} M_\odot$, see (\ref{Vflat_vs_M}).
We take $(dn_\textrm{sim}/dV_\textrm{flat})/(dn_\textrm{obs}/dV_\textrm{flat}) \approx 14$
at $V_\textrm{flat} = 20$ km/s, corresponding to $M \approx 10^{9.2} M_\odot$.

We proceed as follows for each of the panels in Figures \ref{z8_Dzz}, \ref{z6_Dzz} and \ref{z4_Dzz}.
We shift the $\Lambda$CDM prediction to the left until agreement with the
data is obtained at $\log_{10}(x) = 10.9$,
where $x$ is $M/M_\odot$, or $10^{1.5} M_*/M_\odot$, or $\nu L_\textrm{UV}/L_\odot$.
We then follow the shifted $\Lambda$CDM prediction to $\log_{10}(x) = 9.2$,
and compare with the data.
If the corresponding ratio is in the approximate range 14 to 7 (to account
for satellites found since the publication of \citet{Klypin}), and a good
fit is obtained with $k_\textrm{fs}(t_\textrm{eq}) = 2.0^{+0.8}_{-0.5} \textrm{ Mpc}^{-1}$ \citep{UVL}, we regard the 
parameter $n$ of the prediction to be ``good". If there is some tension, we clasify $n$ as ``fair".
A summary is presented in Table \ref{quality}.
We conclude that for $0.3 \lesssim n \lesssim 0.8$, the predicted and observed
``Missing Satellites" are in agreement, for galaxies formed with redshift 
$z \gtrsim 6$.

\begin{table}
\begin{center}
{\small
\caption{\label{quality}
Values of the non-linear small scale regeneration parameter $n$
that obtain ``good", ``fair", or ``poor" agreement with the
``Missing Satellites Problem", as a function of	the redshift of galaxy formation $z$,
obtained from the panels in Figures \ref{z8_Dzz}, \ref{z6_Dzz}, and \ref{z4_Dzz}.
}
\begin{tabular}{lccl}
\hline
\hline
Redshift $z$ & Good & Fair & Poor \\
\hline
8 & 0.3, 0.5, 0.7 &               & 0.1, 1.1, 2.0 \\
6 & 0.3, 0.5, 0.7 &               & 0.1, 1.1, 2.0 \\
4 & 	          & 0.7 & 0.1, 0.3, 0.5, 1.1, 2.0 \\
\hline
\hline
\end{tabular}
}
\end{center}
\end{table}

\section{The UV Luminosity Cut-Off}

Reionization begins in earnest at $z \approx 8$, and ends at $z \approx 6$.
For each panel of Figure \ref{z8_Dzz}, corresponding to $z = 8$, we integrate numerically the UV luminosity
along the appropriate ellipsoidal collapse prediction (with parameter $0.84 \nu$),
that obtains excellent agreement with the data.
The following procedure is followed in \citet{Lapi2}: the observed UV luminosity
distribution is extended (without the $\Delta z$ velocity dispersion cut-off)
to an assumed sharp UV magnitude cut-off $M_\textrm{UV}$, and the corresponding reionization
optical depth $\tau$ is calculated. Here we obtain the equivalent sharp UV magnitude cut-off $M_\textrm{UV}$, and
the corresponding reionization optical depth $\tau$ from \citet{Lapi2}.
The results are summarized in Table \ref{tau}.
We note	that, for the range $0.5 \lesssim n \lesssim 0.8$, we obtain agreement with
the measured reionization optical depth	$\tau =	0.053 \pm 0.007$ obtained by the
Planck collaboration \citep{Planck} \citep{PDG2022}.

\begin{table}
\begin{center}
\caption{\label{tau}
For each $n$
we obtain the equivalent sharp UV magnitude cut-off 
$M_\textrm{UV} \approx 5.9 - 2.5 \log_{10}{(\nu L_\textrm{UV}/L_\odot)}$, and the
corresponding reionization optical depth $\tau$ from \citet{Lapi2}.
For comparison, the Planck collaboration measurement is $\tau = 0.053 \pm 0.007$
\citep{Planck} \citep{PDG2022}.
}
\begin{tabular}{lcccr}
\hline
\hline
$n$ & $M_\textrm{UV}$        & $\tau$ & fit quality \\
\hline
0.9 & -18.6 & $0.050 \pm 0.006$ & fair      \\
0.8 & -18.3 & $0.050 \pm 0.006$ & good      \\
0.7 & -17.6 & $0.052 \pm 0.006$ & excellent \\
0.6 & -16.9 & $0.053 \pm 0.008$ & excellent \\
0.5 & -14.9 & $0.059 \pm 0.008$ & excellent \\
0.4 & $>-11.9$ & $> 0.07$          & good      \\
0.3 & $>-11.9$ & $> 0.07$          & fair      \\
\hline
\hline
\end{tabular}
\end{center}
\end{table}

\section{Conclusions}

Comparisons of galaxy UV luminosity distributions, and galaxy stellar mass  
distributions, with predictions for $M > M_\textrm{vd}$, obtain
the free-streaming cut-off wavenumber
$k_\textrm{fs}(t_\textrm{eq}) =	2.0^{+0.8}_{-0.5} \textrm{ Mpc}^{-1}$,
with the non-linear regeneration of small scale	structure parameter $n$ in
the wide approximate range 0.2 to 1.1 \citep{UVL}.
In the present work we have extended the predictions to $M < M_\textrm{vd}$,
including the free-streaming cut-off (\ref{tail}), and the velocity dispersion cut-off
of Table 2.
This extension is in quantitative agreement with the ``Missing Satellites Problem"
for galaxies formed at $z \gtrsim 6$,
and with the needed UV cut-off (to not exceed the observed reionization
optical	depth), with $n$ in the approximate range 0.5 to 0.8.

As a cross-check, we have obtained the adiabatic invariant in the core of
dwarf galaxies dominated by dark matter, from their rotation curves.
The result is $v_{h\textrm{rms}}(1) = 0.406 \pm 0.069$ km/s \citep{dwarf}, corresponding to
a free-streaming cut-off wavenumber
$k_\textrm{fs}(t_\textrm{eq}) = 1.90 \pm 0.32 \textrm{ Mpc}^{-1}$,
from Equation (\ref{kfs}). 
The agreement of these two independent measurements of $k_\textrm{fs}(t_\textrm{eq})$
confirms 1) that
the adiabatic invariant	in the core of galaxies	is of cosmological origin,
as predicted for warm dark matter \citep{first_galaxies},
since several galaxies accurately share the same adiabatic invariant,
and 2) confirms	that $k_\textrm{fs}(t_\textrm{eq})$ is due to free-streaming.
All of these results are data driven.

As a by-product of this study we obtain the empirical Tully-Fisher relation
from first principles, by integrating numerically the galaxy formation
hydro-dynamical equations \citep{first_galaxies}. These hydro-dynamical equations predict that
the core of first galaxies form adiabatically if dark matter is warm, i.e. conserves
$v_{h\textrm{rms}}(1)$.

Omitting the non-linear	regeneration of	small scale structure, i.e. setting $n = 2$,
or using the similar $\tau^2(k)$ from the linear Equation (7) of \citet{Viel2013}, 
and omitting the velocity dispersion cut-off,
obtains	strong disagreement with observations. These omissions have led several
published studies to obtain lower warm dark matter particle ``thermal relic mass" limits
of several keV.
Note that nature, and simulations \citep{White}, re-generate non-linear small scale
structure when relative density perturbations approach unity.
May I suggest that these limits	be revised, including the 
non-linear regeneration of small scale structure, and the velocity dispersion cut-off.
We note	that the Particle Data Group's ``Review of Particle Physics (2022)"
quotes lower limits of 70 eV for fermion dark matter, or $10^{-22}$ eV for
bosons \citep{PDG2022}, not several keV.

To summarize, warm dark matter with	
an adiabatic invariant $v_{h\textrm{rms}}(1) = 0.406 \pm 0.069 \textrm{ km/s}$ \citep{dwarf},
a free-streaming comoving cut-off wavenumber
$k_\textrm{fs}(t_\textrm{eq}) = 2.0^{+0.8}_{-0.5} \textrm{ Mpc}^{-1}$ \citep{UVL},
and a non-linear small scale regenerated ``tail"
as in (\ref{tail}) with $0.5 \lesssim n \lesssim 0.8$, 
is in agreement with galaxy rotation curves \citep{dwarf}, galaxy
stellar	mass distributions, galaxy rest	frame UV luminosity 
distributions \citep{UVL}, the Missing Satellites Problem,
and the	UV luminosity cut-off needed to	not exceed the
measured reionization optical depth.           \\

\textbf{Data availability:} \citep{So} \citep{Gr} \citep{Da} \citep{Bouwens}
\citep{Bouwens2021} \citep{McLeod} \citep{Bouwens_27}

%%%%%%%%%%%%%%%%%%%%%%%%%%%%%%%%%%%%%%%%%%%%%%%%%%

% Don't change these lines
\bsp    % typesetting comment
\label{lastpage}

\begin{thebibliography}{99}

%\bibitem[\protect\citeauthoryear{Author}{2013}]{author2013}
%\bibitem[\protect\citeauthoryear{Jones}{2015}]{jones2015}
%Jones C.~D., 2015, Journal of Interesting Stuff, 17, 198
%\bibitem[\protect\citeauthoryear{Smith}{2014}]{smith2014}
%Smith A.~B., 2014, The Example Journal, 12, 345 (Paper I)

\bibitem[\protect\citeauthoryear{Bouwens}{2014}]{Bouwens_27}
Bouwens R.~J., Illingworth G.~D., Oesch P.~A., 2014,
Astrophysical Journal, 793, 115

\bibitem[\protect\citeauthoryear{Bouwens}{2015}]{Bouwens}
Bouwens R.~J. et al., 2015,
Astrophysical Journal, 803, 34

\bibitem[\protect\citeauthoryear{Bouwens}{2021}]{Bouwens2021}
Bouwens R.~J. et al., 2021,
Astronomical Journal, 162, 2 

\bibitem[\protect\citeauthoryear{Boyanovsky}{2008}]{Boyanovsky}
Boyanovsky D., de  Vega H.~J., Sanchez N.~G., 2008,
Physical Review D, 78, ID:063546

\bibitem[\protect\citeauthoryear{Davidzon}{2017}]{Da}
Davidzon I.,  Ilbert O.,  Laigle C. et al., 2017,
Astronomy and Astrophysics, 605, idA70

\bibitem[\protect\citeauthoryear{Grazian}{2015}]{Gr}
Grazian A. et al., 2015,
Astronomy and Astrophysics, 575, A96

\bibitem[\protect\citeauthoryear{Hoeneisen}{2022a}]{wdm_measurements_and_limits}
Hoeneisen B., 2022a
International Journal of Astronomy and Astrophysics, 12, 94

\bibitem[\protect\citeauthoryear{Hoeneisen}{2022b}]{first_galaxies}
Hoeneisen B., 2022b,
Journal of Modern Physics, 13, 932

\bibitem[\protect\citeauthoryear{Hoeneisen}{2022c}]{UVL}
Hoeneisen B., 2022c,
International Journal of Astronomy and Astrophysics, 12, 258

\bibitem[\protect\citeauthoryear{Hoeneisen}{2022d}]{dwarf}
Hoeneisen B., 2022d,
International Journal of Astronomy and Astrophysics, 12, xxx

\bibitem[\protect\citeauthoryear{Klypin}{1999}]{Klypin}
Klypin A.~A., Kravstov A.~V., Valenzuela O., 1999,
Astrophysical Journal, 522, 82

\bibitem[\protect\citeauthoryear{Lapi}{2015}]{Lapi2}
Lapi A., Danese L., 2015, Journal of Cosmology and Astroparticle Physics, 9, no:3

\bibitem[\protect\citeauthoryear{Lapi}{2017}]{Lapi_SMF}
Lapi A. et al., 2017,
Astrophysical Journal, 847, no:13

\bibitem[\protect\citeauthoryear{Markovi\u{c}}{2013}]{Viel2013}
Markovi\u{c}, Viel M., 2013,
Cambridge University Press, Cambridge

\bibitem[\protect\citeauthoryear{Mason}{2015}]{Mason}
Mason C.~A., Trenti M., Treu T., 2015,
Astrophysical Journal, 813, 21

\bibitem[\protect\citeauthoryear{McLeod}{2015}]{McLeod}
McLeod D.~J. et al., 2015,
MNRAS 450, 3032

\bibitem[\protect\citeauthoryear{Paduroiu}{2015}]{Pfenniger}
Paduroiu S., Revaz Y., Pfenniger D., 2015,
https://arxiv.org/pdf/1506.03789.pdf

\bibitem[\protect\citeauthoryear{Planck}{2018}]{Planck}
Planck Collaboration Results VI (2018), [arXiv:1807.06209]

\bibitem[\protect\citeauthoryear{Press-Schechter}{1974}]{Press-Schechter}
Press W.~H., and Schechter P., 1974,
Astrophysical Journal, 187, 425

\bibitem[\protect\citeauthoryear{Sheth-Mo-Tormen}{2001}]{Sheth_Mo_Tormen}
Sheth R.~K., Mo H.~J., Tormen G., 2001,
MNRAS, 323, 1

\bibitem[\protect\citeauthoryear{Sheth-Tormen}{1999}]{Sheth_Tormen}
Sheth R.~K., Tormen G., 1999,
MNRAS, 308, 119

\bibitem[\protect\citeauthoryear{Song}{2016}]{So}
Song M., Finkelstein S. L., Ashby M. L. N., et al., 2016,
Astrophysical Journal, 825, 5

\bibitem[\protect\citeauthoryear{Weinberg}{2008}]{Weinberg}
Weinberg S., 2008, Cosmology,
Oxford University Press, Oxford OX2 6DP

\bibitem[\protect\citeauthoryear{White}{2018}]{White}
White M., Croft R.~A.~C., 2018,
Astrophysical Journal, 539, 497

\bibitem[\protect\citeauthoryear{Workman}{2022}]{PDG2022}
Workman R.~L. et al., 2022 (Particle Data Group),
The Review of Particle Physics (2022)
Prog. Theor. Exp. Phys. 083C01

\end{thebibliography}
\end{document}